\documentclass[journal]{IEEEtran}
\def\ps@headings{%
\def\@oddhead{\mbox{}\scriptsize\rightmark \hfil \thepage}%
\def\@evenhead{\scriptsize\thepage \hfil \leftmark\mbox{}}%
\def\@oddfoot{}%
\def\@evenfoot{}}
\makeatother
\usepackage{booktabs} 
\usepackage{url}
\usepackage{svg}
\usepackage{etoolbox}
\usepackage{cuted}
\usepackage{lipsum}
\usepackage{amssymb}
\usepackage{array}
\usepackage{amsmath}
\usepackage{tikz}
\usepackage{color}
\usepackage{dirtytalk}
\usepackage{algpseudocode}
\usepackage{graphicx}
\usepackage{caption}
\usepackage{mdframed}
\usepackage{tcolorbox}
\usepackage{comment}
\usepackage[english]{babel}
\usepackage[utf8x]{inputenc}
\usepackage{amsmath,amsthm}
\usepackage{multirow}
\usepackage[ruled,vlined]{algorithm2e}
\usepackage{textcomp}
\usepackage{booktabs,makecell}
    
\usepackage{siunitx}

\theoremstyle{definition}

\theoremstyle{plain}

\ifCLASSOPTIONcompsoc
  \usepackage[nocompress]{cite}
\else
  \usepackage{cite}
\fi
\ifCLASSINFOpdf
\else
\fi
\begin{document}
%

\title{{A~Hybrid Blockchain-Edge Architecture for Electronic~Health~Record Management with  Attribute-based Cryptographic Mechanisms} 
}

\author{Hao Guo,~\IEEEmembership{Member,~IEEE,}
        Wanxin Li*,~\IEEEmembership{Member,~IEEE,}
        Mark Nejad,~\IEEEmembership{Member,~IEEE,} \\
        and~Chien-Chung Shen,~\IEEEmembership{Member,~IEEE}
\thanks{Manuscript received November 21, 2021; revised March 24, 2022 and June 13, 2022, accepted June 20, 2022. \textit{(Corresponding author: Wanxin Li.)}}
\IEEEcompsocitemizethanks{
\IEEEcompsocthanksitem Hao Guo is with the Research \& Development Institute of Northwestern Polytechnical University in Shenzhen, 518057, and the School of Software, Northwestern Polytechnical University, Xi'an, 710129, China (e-mail: haoguo@nwpu.edu.cn).
\IEEEcompsocthanksitem Wanxin Li is with the Department of Communications and Networking, Xi'an Jiaotong-Liverpool University, Suzhou, 215123, China, and the Department of Civil and Environmental Engineering, University of Delaware, Newark, Delaware, 19716, USA (e-mail: wanxinli@udel.edu).
\IEEEcompsocthanksitem Mark Nejad are with the Department of Civil and Environmental Engineering, University of Delaware, Newark, Delaware, 19716, USA (e-mail: nejad@udel.edu).
\IEEEcompsocthanksitem Chien-Chung Shen is with the Department of Computer and Information Sciences, University of Delaware, Newark, Delaware, 19716, USA (e-mail: cshen@udel.edu).
}
}


%
%

\markboth{IEEE Transactions on Network and Service Management
,~Vol.~XX, No.~XX,~2022}%
{Shell \MakeLowercase{\textit{et al.}}: Bare Demo of IEEEtran.cls for Computer Society Journals}
\maketitle
\pagestyle{headings} 

\begin{abstract}

This paper presents a hybrid blockchain-edge architecture for managing Electronic Health Records (EHRs) with attribute-based cryptographic mechanisms. The architecture introduces a novel attribute-based signature aggregation (ABSA) scheme and multi-authority attribute-based encryption (MA-ABE) integrated with Paillier homomorphic encryption (HE) to protect patients' anonymity and safeguard their EHRs. All the EHR activities and access control events are recorded permanently as blockchain transactions.
We develop the ABSA module on Hyperledger Ursa cryptography library, MA-ABE module on OpenABE toolset, and blockchain network on Hyperledger Fabric. 
We measure the execution time of ABSA's signing and verification functions, MA-ABE with different access policies and homomorphic encryption schemes, and compare the results with other existing blockchain-based EHR systems. We validate the access activities and authentication events recorded in blockchain transactions and evaluate the transaction throughput and latency using Hyperledger Caliper.
The results show that the performance meets real-world scenarios' requirements while safeguarding EHR and is robust against unauthorized retrievals.

\end{abstract}
\IEEEpeerreviewmaketitle

\section{Introduction}\label{sec:introduction}

%
%
%
%


\IEEEPARstart 
{I}{n} healthcare, electronic health records (EHRs) contain highly sensitive personal data for the diagnosis, treatment, and management of patients, which are regularly updated, accessed, and shared by multiple parties including doctors, nurses, hospitals, pharmacies, medical researchers, and insurance companies \cite{dubovitskaya2017secure}. This poses a major challenge to EHR management for storing, updating, and sharing EHRs without compromising their security or violating patients' privacy. 
For instance, according to the U.S. Department of Health and Human Services' Office for Civil Rights (OCR), the largest Health Insurance Portability and Accountability Act (HIPAA) violation penalty of 2020 was imposed on the health insurer Premera Blue Cross for a data breach of 10,466,692 EHR records~\cite{hipaajournal}.

{To address this challenge, the US HIPAA sets guidelines to modernize and protect the flow of electronic healthcare information. In a typical EHR access scenario, although healthcare providers (termed data users), such as doctors, nurses, medical researchers, and pathologists, need to authenticate the patients, not all of them (for example, pathologists) need to access all the private EHR data.}
To authenticate patients' identities while protecting their privacy, the mechanisms of Attribute-based Signature (ABS)~\cite{maji2008attribute} and its variants have promising potentials. 
Using ABS, the signature generated and signed with a patient's attributes is attested not to the patient's specific identity information but instead to different attributes owned by the patient. However, existing ABS schemes are not flexible for updating embedded predicates of attributes within the signature since signatures are created through a one-time generation process. Moreover, the verification keys and the size of ABS signatures are large, and  the signature generation and signature verification processes are usually time-consuming and compute-intensive.

To secure the EHR data while facilitating various access policies and sharing mechanisms among different data users, the mechanisms of Attribute-based Encryption (ABE)~\cite{sahai2005fuzzy} and its variants could serve as a potential solution. In general, ABE is a public-key encryption mechanism that can bind security directly between EHR data and data users who want to access them. In particular, CP-ABE \cite{bethencourt2007ciphertext} schemes encrypt EHR which can be accessed by data users with attributes that satisfy the attribute-based access control policies.

By adopting both ABE and ABS schemes for EHR management, patients could define how their EHRs are shared via access control policies while being authenticated without revealing any sensitive identity information.
For example, patient Annie possesses the driver's license issued by the Department of Motor Vehicles (DMV) of Delaware and the insurance card issued by the Blue Safeguard insurance company. With ABS, the signature could be generated with Annie's attributes of her driver's license number and her insurance ID.
With ABE, Annie could grant access permission of her EHR data to participants who are `Doctor' or `Nurse' working in  `Mercy Infirmary.' 



However, conventional ABE solutions require a centralized and trusted private key generator (PKG) to bootstrap the system and distribute secret keys to different participants (e.g., healthcare providers and patients). Several drawbacks can be raised by having this PKG scheme: First, it is hard to find a fully trusted PKG in the real world, which never gets compromised. Secondly, the centralized PKG system has the key escrow issues; the ownership of participants' data is not controlled by themselves. The PKG could decrypt the data, and it may leak sensitive data for some reason. The above scenario also happened to the traditional ABS scheme, which is also limited by a centralized authority.

The blockchain technology has been proposed as a potential solution for EHR management in recent years~\cite{ekblaw2016case,guo2020icbc,halamka2017potential,griggs2018healthcare}. 
However, to maximize the capability of blockchain-inspired EHR data management systems, all the following problems are still required to be fully considered. First is the privacy concerns of the patients and security problems for the EHR data. Due to blockchain's decentralized and transparent characteristics, any patients' sensitive and private data can not be saved directly in the new blockchain transaction. The second is the storage space for each block. Usually, the storage capacity of blocks in blockchain transactions is very limited to accept EHR information, including large-size medical images. 

EHR data are collected and generated from smart sensors and smart medical devices, which require both computation-intensive and delay-sensitive functionalities. Due to the increasing volume of EHR data, performing all processing and data management tasks in a centralized cloud server is no longer an optimal choice. Edge computing~\cite{satyanarayanan2017emergence,shi2016edge,pan2017future} refers to a platform that provides network service, computation, storage, and other applications on sites close to the data source and computing service to address location awareness, network latency, scalability, data security, and privacy issues~\cite{guolhraft2022}.  Edge nodes are located closer to the generation sources of information to reduce bandwidth usage and network latency compared to the cloud computing scenarios. Moreover, edge computing provides the continuation of operation and service despite desultory connections in cloud computing cases. As a result, the EHR data sharing and management system will benefit from edge computing.

This paper proposes a multi-authority verification key and signature aggregation of ABS termed Attribute-based Signature Aggregation (ABSA). Specifically, EHR access activities and authentication events for all participants (patients, healthcare providers, and insurance companies) are permanently saved in a blockchain transaction, i.e., {\em on-chain}, for future accountability and traceability.  Unlike ACL rules utilized in our earlier study \cite{guo2019access}, we develop the novel ABSA scheme compared against a multi-signature threshold to enable authentication procedures and access control events between healthcare providers and patients. Additionally, the {\em off-chain} edge node with IPFS\footnote{https://github.com/ipfs/ipfs}, which acts as the storage,  saves the EHR information encrypted by the MA-ABE mechanism so that the qualified participants who satisfy attribute-based access policies could decrypt and retrieve EHR information from IPFS without the trusted central authority. We also utilize Paillier homomorphic encryption to perform the computations on encrypted EHR data without accessing it.



This paper makes the following contributions: 
\begin{itemize}

\item We proposed a hybrid blockchain-edge architecture to facilitate EHR management. 
We designed a novel ABSA scheme to verify the short multi-signature via the public key aggregation and signature threshold schemes. We integrated MA-ABE with homomorphic encryption to preserve end-to-end privacy and secure EHR data while maintaining EHR activities and access events as permanent blockchain transactions. 


\item We constructed the on-chain/off-chain hybrid edge architecture with IPFS Merkle DAG structure and smart contract-based access control policies. We proposed the formal proof of system correctness analysis, threat model, and privacy analysis discussions for all cryptographic-related mechanisms.

\item We implemented the ABSA, MA-ABE with homomorphic encryption and blockchain modules for the proposed system. We experimented with the ABSA scheme to measure signing and verification time, and tested MA-ABE with homomorphic encryption performances. All results indicate the system has efficient data encryption/decryption and signature aggregation/verification time. Blockchain transaction throughputs and latency are also feasible in EHR activities.
\end{itemize}

The rest of the paper is organized as follows. Related work is discussed in Section II. Section III describes background knowledge for eHealth sensor platforms and the ABS scheme.  We propose the system architecture in Section IV. Specifically, we describe the structure of multi-authority CP-ABE, Paillier homomorphic encryption, ABSA mechanisms, on-chain/off-chain with IPFS storage architecture, access control policies, and the system workflow. In Section V, we conduct the security analysis, privacy analysis, and threat model discussions. In Section VI, we describe the deployment of the prototype to test ABSA and MA-ABE scheme's functionality and experiment with access activities and authentication events. In Section VII, we test performance evaluation and comparison regarding MA-ABE, HE, ABSA and blockchain network. Section VIII concludes the paper with future work directions.
\section{Related Work}





We review state-of-art research works on blockchain-inspired schemes for EHR management. Guo et al.~\cite{guo2018secure} described the blockchain-inspired MA-ABS mechanism, which presented the MA-ABS scheme with the protection of {\it{N-1}} corrupted authorities cannot succeed in collusion attack. Tang et al.~\cite{tang2019efficient} described one efficient authentication solution (IBS) for
blockchain-inspired EHR management system, and prove the security of their mechanism with the random oracle model.
However, both of them did not provide enough information for the  detailed EHR management operations and blockchain system's actions. Yuan et al.~\cite{yuan2017blockchain} proposed a CP-ABE mechanism with blockchain to address the information security issue. However, their solution still relied on one centralized authority to distribute attribute key pairs which hinder the system security.

Zhuang et al.~\cite{zhuang2020patient}  proposed
the blockchain-inspired patient-centric healthcare data-sharing framework to address the security and privacy challenges, the proposed system ensured the eHealth data provenance and gave patients full control over eHealth data. 
Abdullah et al.~\cite{al2017medibchain} also described the patient-centric EHR management system utilizing the blockchain as the storage platform to protect data privacy. In the proposed mechanism, pseudonymity is achieved by using multiple cryptographic techniques to safeguard data.
However, these schemes do not consider storage limitation of actual block sizes to contain EHR information such as videos and medical images.




Li et al.~\cite{li2020healthchain} described the Healthchain to secure EMRs management and data trading for ehealth systems. They also established the Stackelberg pricing scheme to evaluate EMRs data consumers and providers’ interactions.
Xu et al.~\cite{xu2019healthchain} also proposed Healthchain, a blockchain-inspired privacy-preserving solution for eHealth data management. In their scheme, EHR data are encrypted to perform access control rules. Moreover, the data users could add and revoke authorized participants. However, both schemes do not consider the off-chain storage space to protect the EHR data. Abaid et al.~\cite{abaid2019health} presented the Health Access Broker (HAB), the patient-controlled management of personal health records in the cloud.  HAB introduces the auditing and intrusion-detection mechanism based on ABE.

Abdellatif et al.~\cite{abdellatif2021medge} proposed the MEdge-Chain, which leverages blockchain and edge computing to  exchange medical data efficiently.  They introduce the automated patients monitoring solution at the edge and integrate it with the blockchain architecture to optimize EHR data exchanging among multiple participants.
Egala et al.~\cite{egala2021fortified} described  blockchain-inspired architecture, which provides decentralized EHR and smart-contract-based service automation without hindering system privacy and security. They introduced the decentralized Selective Ring based Access Control (SRAC) scheme  to enhance the proposed solution's security capabilities.
Nguyen et al.~\cite{nguyen2021bedgehealth}
proposed the BEdgeHealth system, which integrates mobile-edge computing and blockchain for data sharing and offloading schemes in the distributed hospital architectures. They also designed the smart contract-based authentication mechanism for user access verification. However, all these schemes do not have the data encryption method to safeguard the system security.


Several other works have been proposed to address IoT applications' security via blockchain and other cryptographic techniques. 
Gao et al.~\cite{gao2021blockchain} utilized the blockchain and SGX-enabled edge computing to secure the IoMT data analysis. The blockchain system authenticates the IoMT devices, and the cloud service provides an access policy management mechanism for IoMT data.
Li et al.~\cite{li2021ehrchain} proposed EHRChain, which is a blockchain-inspired EHR system utilizing the homomorphic and attribute-based  cryptosystem to address security and privacy concerns in the medical industry.  Their proposed CP-ABE scheme is indistinguishable under the chosen plaintext attack.
Putra et al.~\cite{putra2021trust} designed a decentralized attribute-based access control mechanism with an auxiliary Trust and Reputation System (TRS) for IoT authorization. 
However, experiments result for blockchain system performance are missing.

There also have been several attempts to address access control policies on data management with blockchain technology.
Xia et al.~\cite{xia2019secured} described a secure fine-grain access control scheme for outsourced data management. It provides  both write and read functions for outsourced information. They adopted blockchain to provide traceability and visibility to enable access control policies over the data owner's outsourced data.
Ren et al.~\cite{ren2021siledger} proposed the SILedger, which is an open, trusted, and decentralized access control scheme based on blockchain and attribute-based encryption.

{Guo et al. described an MA-ABAC scheme by utilizing Ethereum's smart contract \cite{guo2019multi}, and later proposed the fine-grained AC schemes for EHR management with the hybrid blockchain-edge architecture by developing Hyperledger Fabric access control policies and the off-chain edge node in \cite{guo2019access}. Xu et al.~\cite{xutdsc2021} proposed the secure and flexible EMR sharing system by introducing the dual-policy revocable attribute-based encryption and tamper resistance blockchain technology. Their scheme allows data users to detect any unauthorized manipulation. Gao et al.~\cite{gaotrustaccess2020} proposed a trustworthy secure ciphertext-policy and attribute hiding access control scheme based on blockchain, which utilized the multiplicative homomorphic ElGamal cryptosystem to ensure attribute privacy during authorization validation.}

To the best of our knowledge, this research work is the first effort to integrate the on-chain Hyperledger blockchain system and off-chain edge node with IPFS storage architecture for EHR management. Our proposed solution provides a multi-authority ABSA scheme to facilitate the authentication procedure without revealing any sensitive patient's private information and utilizing the off-chain edge nodes with IPFS  to protect ABE-encrypted EHR information. We conducted the security and privacy analysis with threat model discussions and designed the novel prototype of the EHR management system. We conducted extensive experiments for ABSA, MA-ABE, and HE schemes. Section VII presents the detailed table comparison result with other blockchain-based EHR systems.

\section{Background Knowledge}
This section briefly introduces the eHealth sensor platform, edge node, and attribute-based signature schemes. 
\subsection{eHealth Sensor Platform and Edge Node}

\begin{figure}[t]
\centering
\includegraphics[width=0.487\textwidth]{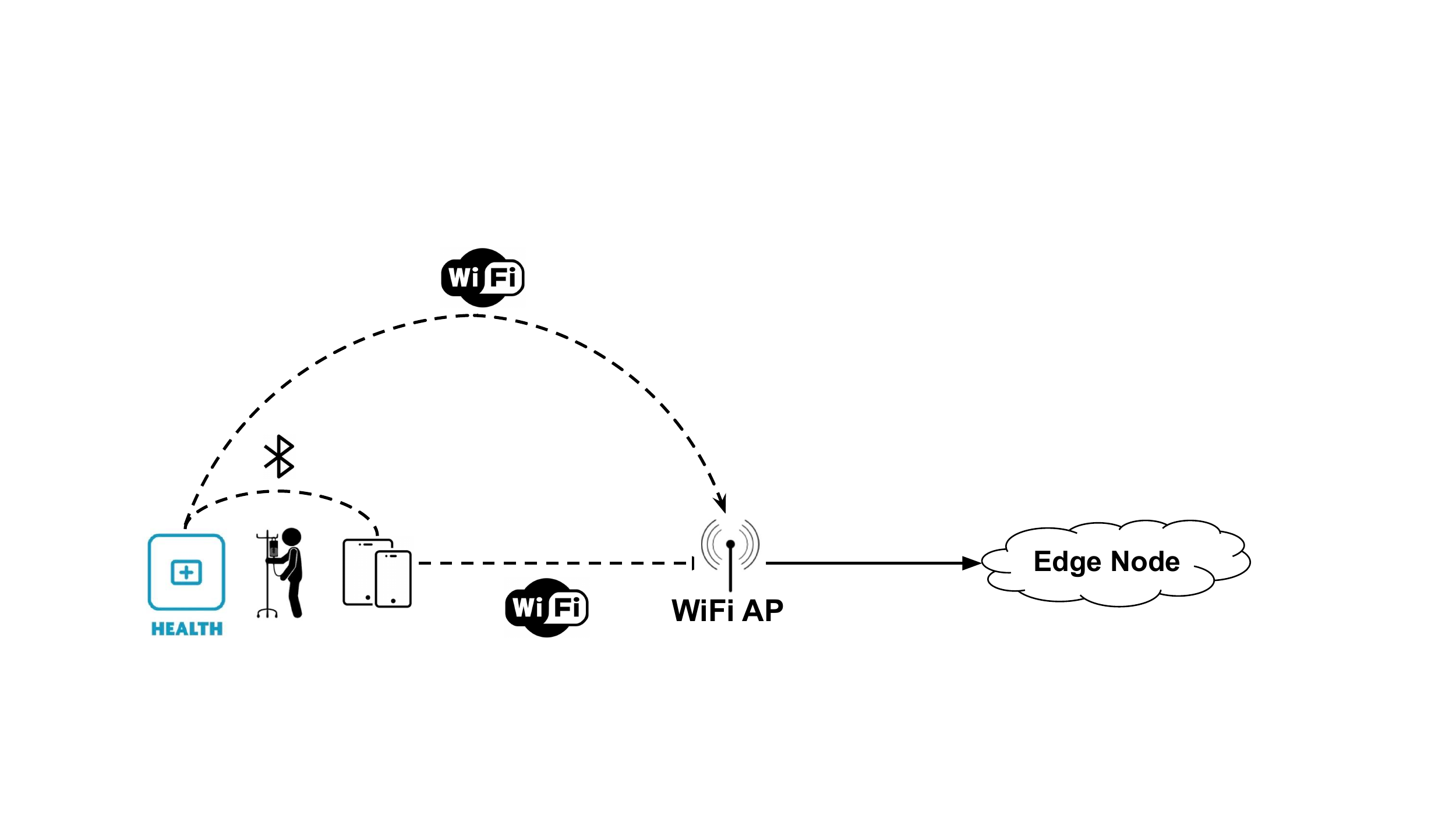}
\caption{EHR collection from smart sensors to edge node.}
\label{fig:sensordata}
\end{figure}

Recent developments in smart wearable sensor technologies empower the EHR management to generate and collect patients' biometric information. For instance, MySignals\footnote{https://www.my-signals.com/.} 
can measure fifteen bio-metric parameters through multiple smart sensors, such as airflow sensor, blood pressure sensor,  Electromyography sensor (EMG), Electrocardiogram sensor (ECG),  Galvanic Skin response sensor, body position sensor, and the snore sensor.
As shown in Fig. \ref{fig:sensordata}, the eHealth sensor device can directly upload the EHR data to the edge storage space through Wi-Fi, or it can first send EHR data to the patient's smart devices through Bluetooth and then uploads it to the edge storage.

\subsection{Attribute-based Signature}

Attribute-based Signature (ABS) is the digital signature scheme that extends from the Identity-based Signature (IBS) scheme~\cite{shamir1984identity} so that the signature is attested to attributes the user possesses instead of a single string to stand for the user's identity. In ABS, users can sign messages
with any predicate of their attributes issued from the attribute authority~\cite{li2010attribute}. 
One signature attests not to the participant's identity who signed the message, but it is a claim regarding attributes to the underlying user possesses~\cite{li2010attribute}. The ABS scheme can protect users' anonymity and privacy by hiding users' identities and corresponding attributes.

According to different policies supported in ABS, threshold-based ABS (t-ABS)~\cite{shahandashti2009threshold,li2010hidden} had been proposed, which supports only a threshold policy under the computational Diffie-Hellman problem. Later, researchers develop a more expressive access policy consisting of AND, OR, and other threshold gates for the ABS mechanism such as~\cite{maji2008attribute} and \cite{cao2011authenticating}. 
By comparison, we propose the ABSA attribute-based signature aggregation mechanism, where each signature attested to attributes is verified independently and then check with the $\it (t, n)$ threshold to provide a more flexible access policy defined by the participants.

\section{System Architecture}
\begin{figure}[t]
\centering
\includegraphics[width=0.48\textwidth]{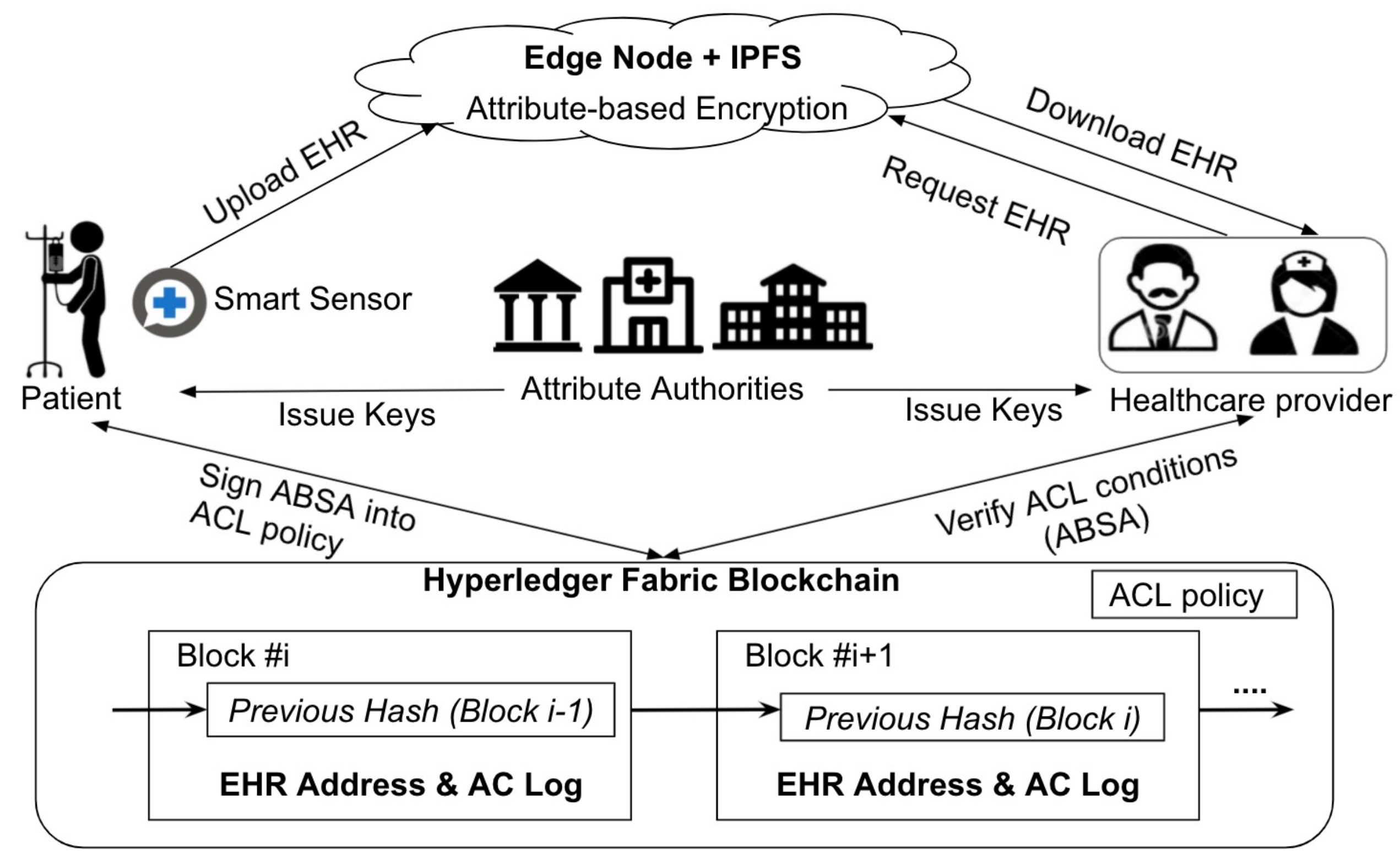}
\caption{EHR Management System Architecture.}
\label{fig:arch}
\end{figure}

This section presents hybrid blockchain-edge architecture composing MA-ABE and ABSA mechanisms to secure patients' EHRs and preserve private identity information. As shown in Fig. \ref{fig:arch}, we describe listed entities.

\begin{itemize}
\item EHR data: EHR data is the information owned by the patient and could be accessed by authorized and qualified healthcare providers who satisfy access control requirements.

\item Patient: A patient is the data owner of their own EHR information; A patient can define access policy for data users (such as doctors and nurses). 

\item Healthcare provider: The healthcare provider (e.g., nurse and doctor) is a data user who wants to access EHR information. The healthcare provider sends the requests for access permissions from the patients (data owner) actively.

\item Attribute: An attribute is the piece of information (e.g., patient's ID) attested to the participants (data owner or the data user). 

\item Attribute authority: An attribute authority is an entity that manages attributes and generates public/private keys to both participants (data owners and users), which is fully trusted.



\item Smart sensor: A smart sensor is a device that collects and generates EHRs from data owners (patients) and then sends it to edge storage, which is semi-trusted.

\item Edge node: An edge node is the storage and computing device, saving EHR information encrypted with the ABE scheme in a remote place, which is semi-trusted.

\item IPFS: The InterPlanetary File System (IPFS) is the peer-to-peer distributed file system. IPFS uses Distributed Hash Table (DHT) and BitSwap technology to establish the peer-to-peer system for robust and fast data storage and block distribution.

\item Smart contracts and blockchain: Smart contracts take patients' signatures as input and return the one-time EHR addresses if the access control policy has been satisfied. The blockchain system acts as the tamper-proof log of authentication events and records the  EHR-related activities for access control events.

\end{itemize}

In the remainder of this section, we first describe multi-authority CP-ABE, HE, and ABSA mechanisms. EHR information presented in the paper is encrypted by multi-authority CP-ABE and homomorphic encryption scheme, whereas the multi-authority ABSA scheme authenticates the 
patients' private attributes (e.g., driver license) anonymously and effectively. The proposed scheme is sufficient for the patient to prove their possession of attributes without revealing sensitive private information. Additionally, we propose on-chain and off-chain hybrid architecture with IPFS storage. We also present detailed construction for access control policies. The workflow of a hybrid blockchain-edge EHR system is presented in the last.






\subsection{Multi-authority CP-ABE Mechanism and Homomorphic Encryption Design}


We present the incorporation of the multi-authority CP-ABE~\cite{lewko2011decentralizing} and homomorphic encryption~\cite{2004Public} scheme in EHR data sharing and access scenarios to avoid the central authority. 
Both data owners and users are granted attribute keys by multiple attribute authorities. Each participant(data owner/ user) in the EHR system gets a global identification number $\it GID$ upon the registration process. Below is detailed construction processes of our proposed scheme:

\textit{Initial Setup}($\lambda$) $\longrightarrow$ $\it IP$. This process takes the security parameter $\lambda$ as input and returns the initial parameter $\it IP$ for the proposed CP-ABE mechanism.

\textit{Authority Setup}($\it IP, AT_i$) $\longrightarrow$ $PK_i$, $SK_i$. Authority $i$ takes  initial parameter $\it IP$ and attribute $AT_i$ as  input value to generate pair of public key $PK_i$ and the secret master key $SK_i$.

\textit{Key Generation}($\it GID, AT_i, {SK_i}, IP$) $\longrightarrow$ $\it {K_i, _{GID}}$. It takes as inputs attribute $\it AT_i$, global identification number $GID$ of participant, secret master key $SK_i$, and initial parameter $\it IP$, and finally outputs attribute key $\it {K_i,_{GID}}$ to participants.

\textit{Encryption}($\it IP, M, \mathbb{A}, \{{PK_i}\}$) $\longrightarrow$ $\it CT$. It takes as inputs initial parameter $\it IP$, EHR information $M$, access policy $\mathbb{A}$ defined by the patient, and public key sets ${PK_i}$. This algorithm will output the ciphertext $\it CT$ as result.

\textit{Decryption} ($\it IP, CT, \{K_i,_{GID}\}$) $\longrightarrow$ $\it M$. It takes as inputs initial parameter $\it IP$, ciphertext $CT$, and collections of attribute keys $\{K_i,_{GID}\}$. This algorithm will return the EHR $\it M$ if and only if the attributes keys set satisfies access control policies imposed on ciphertext $\it CT$. If not, this process will return the failure.

Additionally, we utilize homomorphic encryption~\cite{rivest1978data} to realize security claims and protect the privacy of EHR ciphertext. A homomorphic encryption scheme enables users to perform the computations on its encrypted ciphertext without decrypting it. Predictive analytics in healthcare is difficult to apply through the third-party service provider due to the EHR data privacy concerns. Without accessing the EHR plaintext, insurance companies could realize the security claim without tampering with any patient's privacy by utilizing the Paillier~\cite{2004Public} homomorphic encryption scheme.

{We say that a homomorphic encryption scheme is additive if: 
$Enc(x + y) = Enc(x) +Enc(x)$, and the homomorphic encryption is multiplicative if: $Enc(x \times y) = Enc(x) \times Enc(x)$. Insurance company can utilize the MA-ABE and homomorphic encryption scheme to perform the predictive analytics in ciphertext for privacy-preserving requirements. Paillier’s homomorphic encryption complexity stems from the exponentiation of base $g$. Choosing a small $g$ will expedite computations process drastically. Computation of $r^n$ mod $n^2$ can be computed in key generation procedure, the encryption takes 2 mod $n^2$ power operations and 1 mod $n^2$ multiplication operations, and the decryption
takes $O(|n|^{2+\epsilon})$ bit operations.}

\subsection{Multi-authority ABSA Mechanism Design}

In this subsection, we propose the multi-authority ABSA mechanism for patient authentication to avoid the central authority. In ABSA, a patient's signatures based on their unique attributes are aggregated into one signature and verified through the verification algorithm by applying the public key aggregation scheme and multi-signature threshold scheme. The attribute-based signatures hide the sensitive identity information of the data owner (patients) with corresponding possessed attributes. We improved the BLS signature scheme \cite{boneh2003survey} and the Schnorr signature~\cite{schnorr1991efficient} for the public key aggregation process and signature threshold schemes. The verifier only needs a short multi-signature, and a short aggregation of public keys to conduct the verification procedure. Similar as the BLS construction, our proposed scheme also needs a bilinear pairing  $e$: $\mathbb{G}_1\times \mathbb{G}_2 \rightarrow \mathbb{G}_T$, the hash function $H_0:M \rightarrow \mathbb{G}_1$, and another hash function $H_1:\mathbb{G}^{n}_2  \rightarrow R^n$ where $R:= \{{1, 2,...,2^{128}}\}$.
The high-level construction of the ABSA scheme is shown below:

\textit{Initial Setup}($\lambda$) $\longrightarrow$ $\it IP$: This algorithm takes security parameter $\lambda$ as the input and returns initial parameters $\it IP$ for our proposed scheme.

\textit{Authority Setup}($\it IP, AT_i$) $\longrightarrow$ $\it {SIK_i}, {VK_i}$: Attribute authority runs setup process with initial parameter $\it IP$ and attribute $AT_i$ as inputs, while this algorithm outputs signature key $\it SIK_i$ and  public verification key $\it VK_i$ for every attribute $\it AT_i$.

\textit{Extract}($\it IP, GID, AT_i, {SIK_i}$) $\longrightarrow$ $\it {SK_i, _{GID}}$: This process is executed by multiple attribute authorities. One attribute authority takes initial parameters $\it IP$, participant's $\it GID$, attribute $AT_i$, and authority's signature key ${SIK_i}$ as inputs. Finally, this extract process outputs signing key $\it {SK_i,_{GID}}$.

\textit{Sign}($\it H(A_i), IP, SK_i,_{GID}$) $\longrightarrow$ $\sigma_i$: This process is executed $\it n$ times based on the number of attributes which is belonging to the data owner (patients). This algorithm takes hashed attribute value $\it H(A_i)$, the initial parameter $\it IP$, and the signing key $\it {SK_i,_{GID}}$ as inputs. Eventually, it returns signature $\sigma_i$ as the result.

\textit{Signature Aggregation}($\sigma_1, \sigma_2,..., \sigma_t \longrightarrow \sigma_A $): This algorithm first takes all the individual signatures related to different attributes $\sigma_i$ and compute $(t_1, t_2,...,t_n) \longleftarrow H_1(VK_1, VK_2,...,VK_n)$ and outputs the aggregation signature $\sigma_A \longleftarrow \sigma^{t1}_1\cdot\cdot\cdot \sigma^{tn}_n$.


\textit{Verification Key Aggregation}($\it VK_1, \it VK_2,...,\it VK_t \longrightarrow \it VK_{agg} $) where $0\leq i \leq n-t$. This algorithm first takes all the verification keys related to the corresponding attributes $\it VK_i$ and next compute $(t_1, t_2,...,t_n) \longleftarrow H_1(VK_1, VK_2,...,VK_n)$, and then outputs the aggregation result of the verification key $VK_{agg} \longleftarrow H_1(VK^{t1}_1, VK^{t2}_2,...,VK^{tn}_n)$.

\textit{Verify}($\it H(A), IP, \sigma_A, VK_{agg}$) $\longrightarrow$ $\{0,1\}$: It takes hashed attribute value $\it H(A)$,  initial parameter $\it IP$,  aggregated signature $\sigma_A$, and the aggregation of verification keys $VK_{agg}$ to check that  $e(g_1, \sigma_A) = e(\it VK_{agg},  \it H_0(A))$ condition.
At last, the algorithm returns a Boolean result either $\it accept$ or $\it reject$ to determine whether the aggregated signature is true or false without compromising the patient's sensitive information.

\textbf{Construction.} Assuming that the data owner's global identity is $\it GID$, and he had a series of attributes ({\it t}) and defined the $\it (t, n)$ threshold. He signed the hashed attribute value $\it H(A)$ to get the aggregated signature $\sigma_A$. Next, we prove that the aggregated signature $\sigma_A$ can be verified by the \textbf{Verify} algorithm. First, the aggregated signature $\sigma_A$ is computed as: 
\begin{equation}
   \{\it H_0(A), SK_1\}\longrightarrow \sigma_1,
\end{equation}

\begin{equation}
   \{\it H_0(A), SK_2\}\longrightarrow \sigma_2,
\end{equation}

\begin{equation}
   \{\it H_0(A), SK_i\}\longrightarrow \sigma_i,
\end{equation}

 \begin{equation}
   \sigma^{t1}_1, \sigma^{t2}_2,\cdot\cdot\cdot, \sigma^{tn}_n \longrightarrow \sigma_A,
\end{equation}
and then verifying the aggregated signature $\sigma_A$ is done by checking that iff 
 \begin{equation}
   e(g_1, \sigma_A) = e(\it VK_{agg},  \it H_0(A)),
\end{equation}

For detailed steps of how aggregated signature $\sigma_A$ is constructed, we have the following based on signatures aggregation, verification keys aggregation, and Lagrange interpolation:

\begin{multline}
\begin{split}
    \sigma_A & = \sum\limits_{i=1}^{t-1}\sigma_{i}\prod\limits_{j=0, j\neq i}^{t-1}\frac{x_j}{x_j - x_i} \\
   & = \sum\limits_{i=1}^{t-1}(H_0(A_i)^{ SK_{i}})\prod\limits_{j=0, j\neq i}^{t-1}\frac{x_j}{x_j - x_i} \\
   & = (H_0(A)^{\sum\limits_{i=1}^{t-1} SK_{i}}) \cdot\prod\limits_{j=0, j\neq i}^{t-1}\frac{x_j}{x_j - x_i}\\
    & = ((H_0(A)^ {SK_{1}})^{t1}_1)\cdot\cdot\cdot((H_0(A)^ {SK_{n}})^{tn}_1),\\
    \end{split}
\end{multline}




Next, we prove the Equation. 5:
\begin{multline*}
\begin{split}
e(g_1,\sigma_A) & = e(g_1,\sigma_1 +\sigma_2 +\dots \sigma_t) \\  &=e(g_1,\sigma_1)\cdot e(g_1,\sigma_2)\cdot\dots e(g_1,\sigma_t)\\
 &  =e(g_1,SK_1 \times H_0(A))\cdot\dots e(g_1,SK_t \times H_0(A)) \\
  & = e(\it SK_1\times g_1, H_0(A))\cdot\dots e(\it SK_t \times g_1, H_0(A)) \\
  &  = e(\it VK_1, H_0(A))\cdot\dots e(\it VK_t, H_0(A)) \\
    & = e(\it VK_1\cdot VK_2 \dots VK_t,  H_0(A)) \\
  &  = e(\it VK_{agg},  H_0(A)). \\
\end{split}
\end{multline*}


\begin{figure}[t]
\centering
\includegraphics[width=0.488\textwidth]{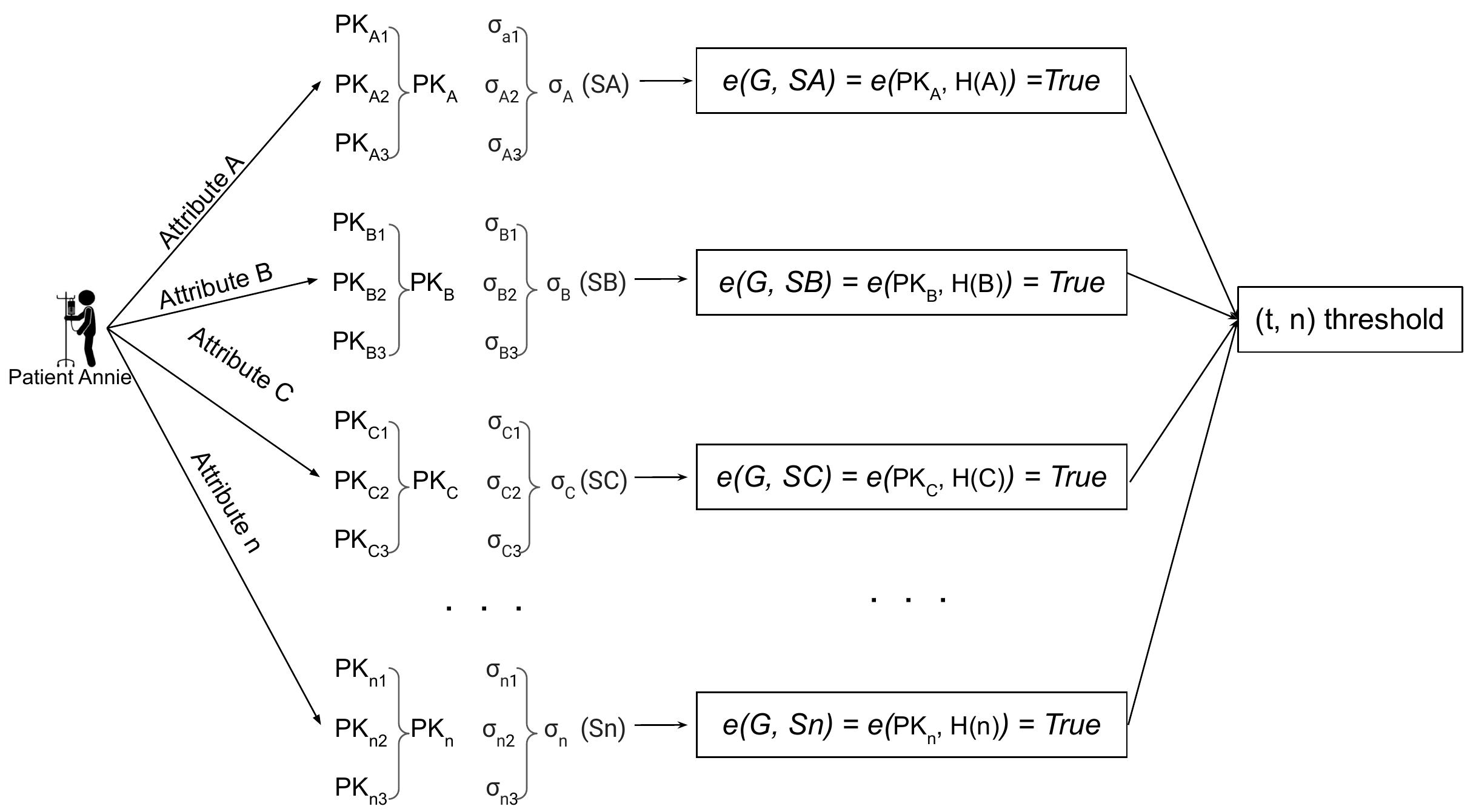}
\caption{t-of-n attribute-based signature aggregation threshold scheme.}
\label{fig:absasig}
\end{figure}

In the proposed ABSA mechanism, the $\it (t, n)$ threshold scheme has been applied based on our proposed public key aggregation and signature aggregation schemes, where $1\leq t \leq n$, as shown in Fig.~\ref{fig:absasig}.
If patient Annie generated three different signatures according to her patient ID issued by Mercy hospital, driver's license issued by DMV, and insurance ID card issued by Blue Cross insurance company. A healthcare researcher, who conducts the routine urine examination for Annie, will follow the threshold of {\em 3 out of 3} to authenticate her identity information. In contrast, a doctor applies the threshold of {\em 1 out of 3} to access the EHR information for regular healthcare analyses. 
Our proposed ABSA aims anonymously to authenticate patients' private attributes for all mentioned scenarios while utilizing multi-signature threshold schemes with more flexibility.

\subsection{On-chain and Off-chain Hybrid Architecture}

\begin{figure}[t]
\centering
\includegraphics[width=0.48\textwidth]{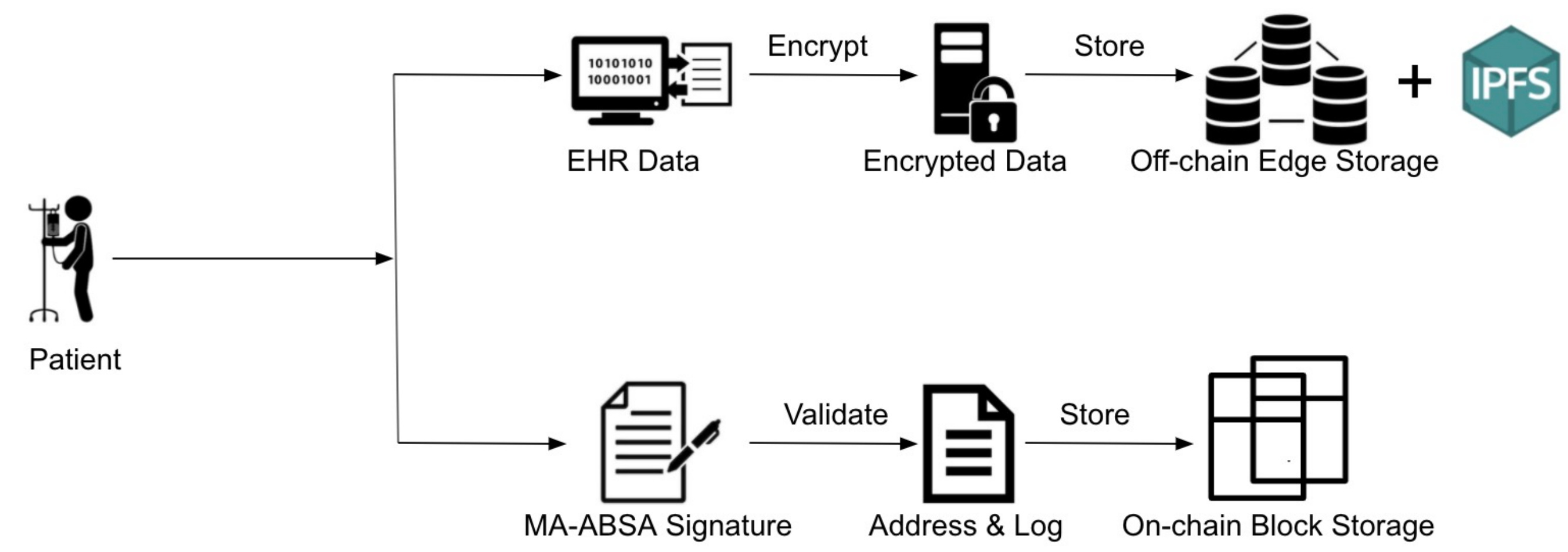}
\caption{Hybrid on-chain and off-chain with IPFS storage architecture.}
\label{fig:onoff}
\end{figure}

As shown in Fig. \ref{fig:onoff}, to address the blockchain transaction space limitation problem~\cite{guo2019access}, the privacy concern of blockchain transaction~\cite{zyskind2015decentralizing}, and the drawback of lacking fine-grained access control policies among multiple users~\cite{guo2019access}. We develop a hybrid blockchain-edge architecture that includes the on-chain mechanism of ABSA authentication events and EHR access activities in the blockchain transaction. Also, the off-chain storage space of ABE-encrypted EHR information in edge storage utilizes the IPFS~\cite{ipfs} distributed file storage system. Blockchain transactions are transparent to each participant where it saves the EHR access logs and addresses. Moreover, sensitive and private patient information is isolated and protected by the proposed hybrid {\it{on-chain/off-chain }} architecture.

Additionally, every patient can store personal information, including name, $\it GID$,  and ABSA signatures in a private storage place. 
The patient can define the attribute-based access policy for data users (e.g., healthcare providers) to access the EHR information, which is implemented with the access control policies of the blockchain system.

We adopted the InterPlanetary File System (IPFS), the peer-to-peer distributed file system, for the off-chain edge storage. IPFS (the InterPlanetary File System) is the hypermedia distribution protocol addressed by identities and content~\cite{ipfs}. IPFS is content-addressable, allocating the unique identifier for every saved data. IPFS applies the Distributed Hash Table (DHT) and the BitSwap to establish a massive P2P system for robust and fast data storage and block distribution~\cite{li2021ehrchain}. 


IPFS builds a Merkle DAG, a
directed acyclic graph where links between objects are cryptographic hashes of the targets embedded in the sources~\cite{dias2016distributed}.
The IPFS Merkle DAG can store data in a flexible way. Every node in a Merkle DAG is the root of a (sub)Merkle DAG itself, and this subgraph is contained in the parent DAG. Merkle DAG nodes are immutable. Any change in a node would alter its identifier and thus affect all the ascendants in the DAG,  and they have self-verified structures.

\subsection{Access Control}

In the proposed architecture, both the patients and doctors act as the participants. The patient’s EHR address is stored privately on the Hyperledger Fabric system. Patients and their EHR assets have unique one-to-one relationships, which are uniquely identified by the patient ID. Authorized by the ACL policy imposed with ABSA signatures and threshold scheme, the doctor can retrieve and check his/her patient’s EHR information while the patient can access their own EHR data only. 
In the last, all historical access and retrieving activities are permanently stored as traceable and immutable EHR access event logs in the Hyperledger blockchain system. 

\begin{figure}[t]
\centering
\includegraphics[width=0.488\textwidth]{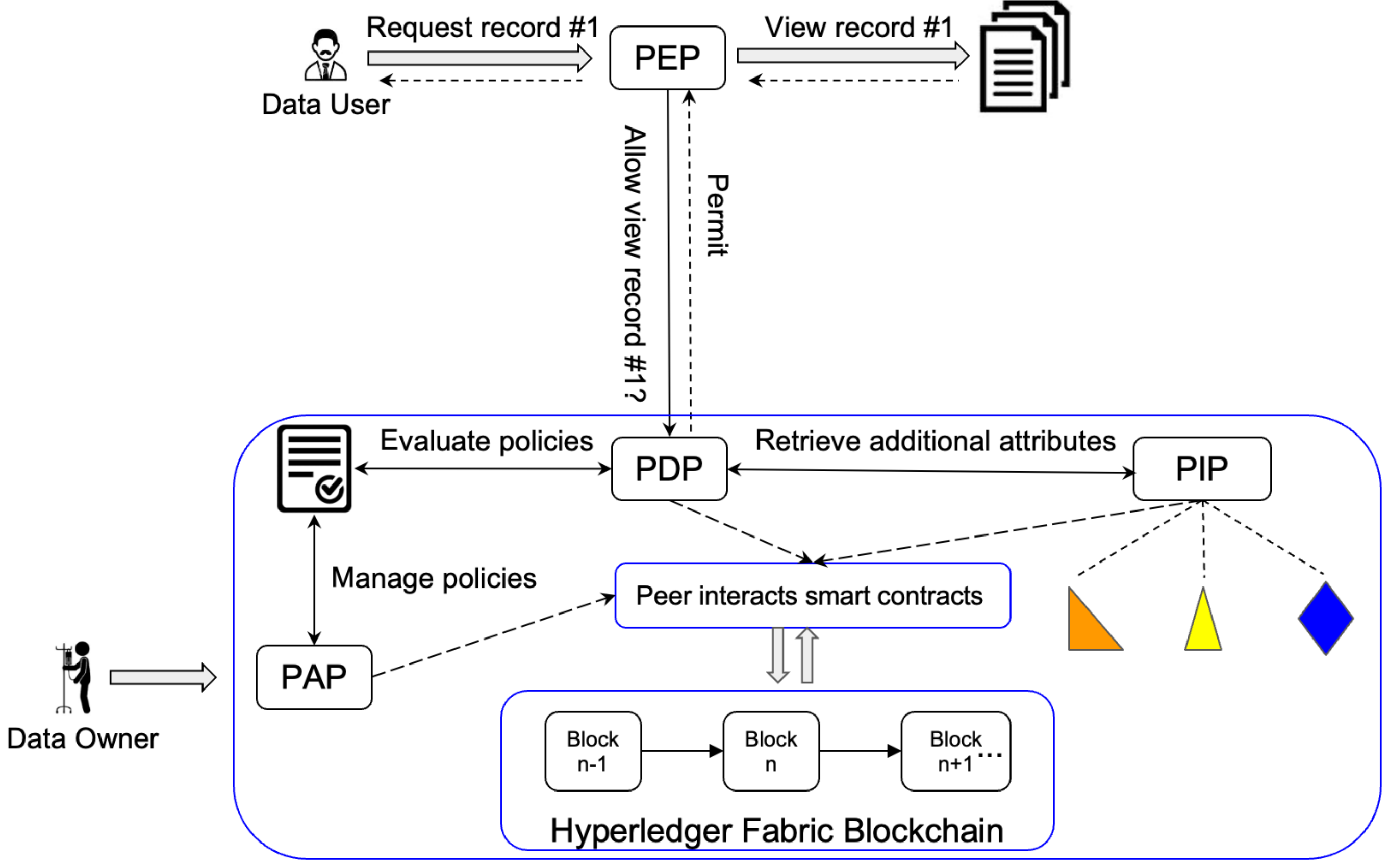}
\caption{Blockchain-inspired Access Control Architecture.}
\label{fig:acblock}
\end{figure}

As shown in Fig.~\ref{fig:acblock}, the blockchain acts as a controller between the data owner and data user, which manages multiple access control components via smart contracts and access event logs on the tamper-proof ledger~\cite{rouhani2020distributed}. Inspired by XACML, we propose three mainly smart contracts for the participants to interact with the policy decision point (PDP), policy information point (PIP), and the policy administration point (PAP) function.

ACL policies are defined with the following components:
\begin{itemize}
\item Subjects: It defines the entity or the person which is participated in the access control process. 

\item Operations: It shows action with governed rules. In the proposed mechanism,  we support three actions: READ, WRITE, and UPDATE. 

\item Objects: It defines objects to which ACL rules apply. In our scenario, it could be either a single document or a complicated union of EHR information. 

\item Conditions: It is an AND-gate access control policies expression over different parameters and  variables. Moreover, the proposed mechanism could support other expressions for more complicated access control policy conditions.

\item Actions: The decision of access control policies. The result can be either ALLOW or DENY action.
\end{itemize}

In our proposed scheme, there are  two kinds of access control policies: non-conditional and conditional. Non-conditional rules are utilized to control ACL policy for the unique group of participants, such as healthcare researchers. On the other hand, conditional ACL rules could indicate different AND-gate ACL policies and output the decision based on the action procedure. For example, {\it Rule1} indicates that only doctors from Mercy Hospital could READ EHR information from the authenticated patients who satisfied the threshold of the ABSA scheme proposed in the previous section:
\begin{verbatim}
rule Rule1 {
  description: "Only doctor from the
   Mercy Hospital could access EHR"
  subject(v): "Mercy.Doctor#102"
  operation: READ
  object(t):"Mercy.patient#205.data"
  condition: "v.role === Doctor &&
   v.organization === Mercy &&
   t.patient_id.verify() === true &&
   t.driver_license.verify() === true &&
   t.insurance_id.verify() === true"
  action: ALLOW
}
\end{verbatim}


According to \cite{2012Designing},  the total
processing time of 100,000 single-valued requests and multi-valued requests respectively for Sun PDP is around 150ms. When the number of rules is 4000 for XACML policies, the total processing time is at the $10^6$ ms level, which indicates the XACML is highly scalable and efficient.

The purpose of the access control mechanism provides the transaction log of access events to EHR information, determining the function for write, read, and update operations. To be more precise, the proposed mechanism provides the functions as to: `who' has the ability to do `what' in our proposed hybrid blockchain-edge framework. 

\begin{figure*}[t]
\centering
\includegraphics[width=0.77\textwidth]{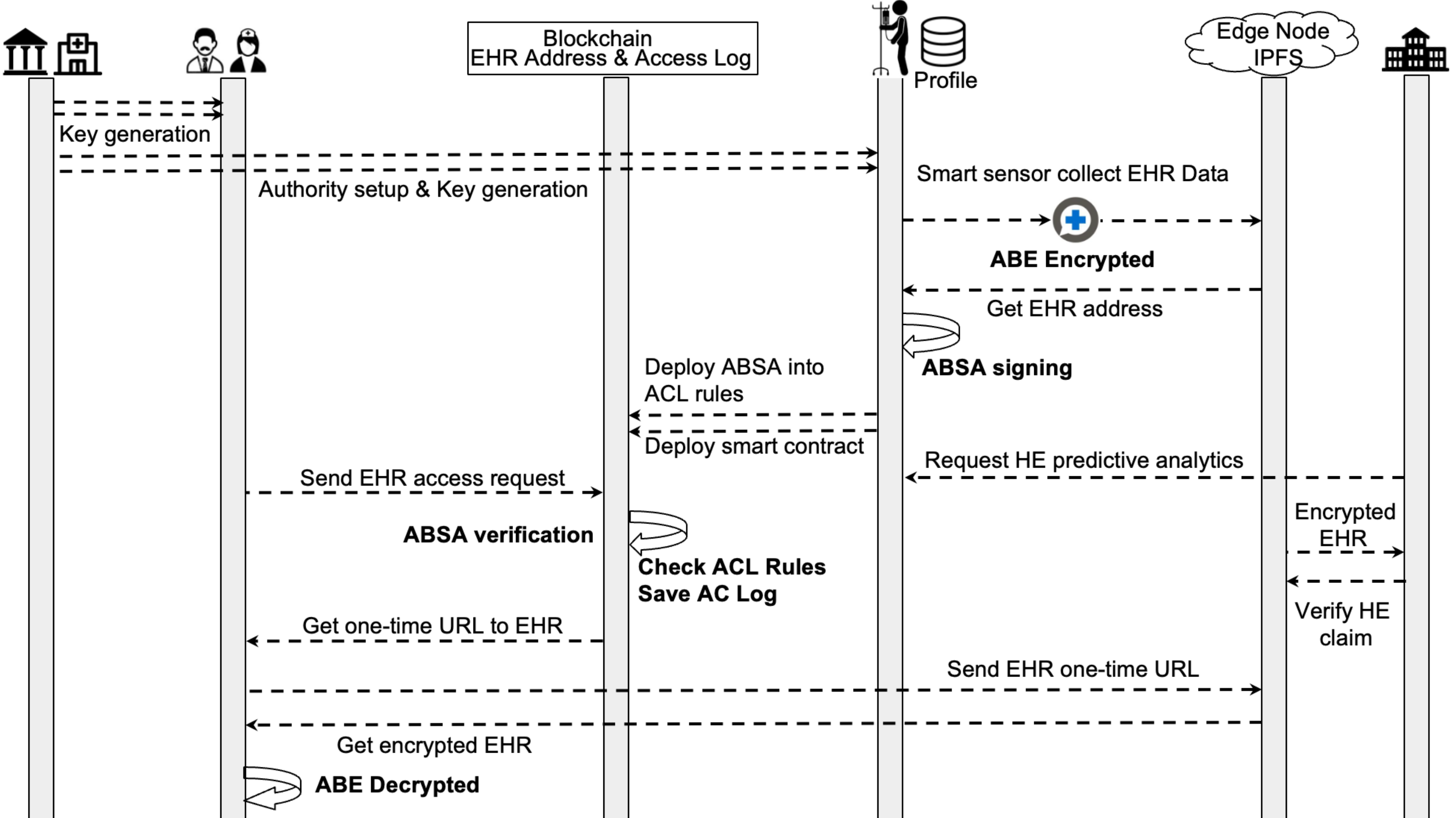}
\caption{Workflow of hybrid blockchain-edge architecture with attribute-based cryptographic mechanisms for EHR management.}
\label{fig:workflow}
\end{figure*}

\subsection{Workflow of EHR Management}


{The workflow of the hybrid blockchain-edge EHR framework is shown in Fig. \ref{fig:workflow}. First, all participants (healthcare providers and patients) register themselves within the proposed system. Next, all attribute authorities will generate ABSA and ABE keys for all participants. Then the patient can upload the EHR information, which is encrypted through the MA-ABE scheme.
Next, the patient can sign the hashed attribute values to generate and aggregate ABSA signatures. 
Next, the participant (doctor) can start the access request for the EHR data. A smart contract is now triggered and executed to save the authentication event into a blockchain transaction and get {\it url} to the doctor. 
Finally, the doctor sends {\it url} to edge with IPFS storage to retrieve encrypted EHR information. When the healthcare provider possesses enough attributes required to satisfy ACL policies pre-defined by the patient, he/she could then decrypt EHR information stored in the edge node.
Additionally, the insurance company could get encrypted EHR from the off-chain IPFS, and utilize the Paillier homomorphic encryption scheme to verify the security claim while keeping the privacy-preserving feature.}

\section{Security and Privacy Analysis with Threat Model Discussions}

\subsection{Security Analysis of ABSA}
Our proposed scheme constructs the ABSA scheme by letting all individual signatures contribute to the multi-signatures on aggregate signatures and public keys during the signature aggregation process. The $i$-th patient in our proposed scheme has the ``membership key," which is the multi-signature on $(VK_{agg},i)$. To verify whether the patient $i$ signed the message, one can check if the aggregated signature is a valid signature where the aggregate verification key of the subgroup signed the message and  membership keys correspond to patient $i$.

\textbf{Signing.} $Sign (SysPar, S, mk_i, sk_i, m)$ computes the $VK_{agg} \leftarrow Kagg(PK)$ and $s_i \leftarrow H_0(VK_{agg}, m)^{sk_i}\cdot mk_i$, where $SysPar$ is the system parameter, $S$ is the subgroup of the aggregate public key, $mk_i$ is ``membership key", $sk_i$ is secret key, and $m$ is attribute message. Next, we compute the
\begin{align*}
    pk \leftarrow \prod\limits_{j\in S} \ pk_j,
\end{align*} \begin{align*}
    s \leftarrow \prod\limits_{j\in S} \ s_j,
\end{align*}
and outputs multi-signature $\sigma : = (pk,s)$. We argue that the subgroup set $S$ can be dynamic and it can be determined when the partial aggregated signatures are collected.

\textbf{Verification.} $Verify (SysPar, vk_{agg}, m, S, \sigma)$ produces $\sigma$ as the $(pk,s)$ and it outputs 1 iff
\begin{align*}
   e(H_0(vk_{agg}, m), pk) \cdot e(\prod\limits_{j\in S}\ H_1(vk_{agg},j),vk_{agg}) \overset{\text{?}}{=} e(s, g_2).
\end{align*}

The proposed ABSA scheme satisfies the correctness. If the participant honestly executes the signature group setup and the signature signing protocols, we then have the \begin{proof}
\begin{align*}
    pk = g_2^{\sum\limits_{i \in S}\ sk_i},
\end{align*} \begin{align*}
    vk_{agg} = g_2^{\prod\limits_{i=1}^{n}\ a_i \cdot pk_i},
\end{align*} and the 
\begin{align*}
  s = H_0{(vk_{agg}, m)^{\sum\limits_{i \in S}\ sk_i}} \cdot
  \prod\limits_{i \in S}^{n} H_1(vk_{agg}, i)^{\sum\limits_{j=1}^{n}\ a_j \cdot sk_j},
\end{align*} which satisfy the ABSA verification process:

\begin{align*}
\begin{split}
    e (s, g_2) & = e(H_0{(vk_{agg}, m)^{\sum\limits_{i \in S}\ sk_i}} \prod\limits_{i \in S}^{n} H_1(vk_{agg}, i)\\^ {\sum\limits_{j=1}^{n}\ a_j \cdot sk_j}, g_2) \\
  & = e(H_0(vk_{agg}, m), pk) \cdot e(\prod\limits_{i \in S}^{n} H_1(vk_{agg}, i), \\  g_2^{\sum\limits_{j=1}^{n}\ a_j \cdot sk_j}) \\
   & = e(H_0(vk_{agg}, m), pk) \cdot e(\prod\limits_{i \in S}^{n} H_1(vk_{agg}, i),\\ vk_{agg}).
  \end{split}
\end{align*}
\end{proof}

Hence we prove the system correctness of the proposed ABSA scheme and pass the verification process.

\subsection{Threat Model}


Our proposed scheme of MA-ABE with Paillier homomorphic encryption enables ciphertext indistinguishability for the EHR ciphertext data stored in IPFS. This can protect the system from the chosen-plaintext attack by the semi-trusted entities such as smart sensors or edge nodes. The security notion for the proposed scheme is defined in the next game.

\textbf{Initialization}. The adversary states data user's attribute set $t$ = ($t_1$,...,$t_k$).

\textbf{Setup}. Next, the challenger executes the $Setup$ algorithm and obtains the public key $pk$ and the secret master key $SK$. The challenger later sends public key $pk$ to the adversary.

\textbf{Phase One}. Now the adversary can issue queries for private keys. Challenger then executes the $Key Generation$ algorithm and returns the $t_k$ to the adversary.

\textbf{Challenge}. An adversary then submits two messages $M_1$ and $M_2$ with equal length, and the adversary provides a pair of access control policies $\big\{$$\mathbb{A}\it{1}$, $\mathbb{A}\it{2}$$\big\}$ that need to be challenged. The condition is that attribute sets here accessed by an adversary cannot satisfy these access control structures. The challenger will randomly choose the $\beta \in (0,1)$, calculate the challenge ciphertext $CT*$ based on these access control policies, and finally return the  $CT*$ to the adversary.

\textbf{Guess}. An adversary outputs a guess $\beta'$ of the $\beta$, and the advantage of an adversary is defined as $Pr$[$\beta'$ = $\beta$] $-$ $1/2$. 

\subsection{Security Claim}





\textbf{Proposition 1.} {\it{ In the real-world predictive analytics process, the challenger (insurance company) verifies whether the security claim needs to be settled.}}
\begin{proof}
\begin{align*}
    \begin{split}
 & Enc(M_a) \times Enc(M)^{-1}  \\
      & = ((1+m_{a}n)r_1^{n}) \times ((1+mn)r^{n})^{-1}~mod~n^2 \\
      & = ((1+\frac{n(m_a - m)}{1+ mn}) \times (\frac{r_1}{r})^{n}~mod~n^2.
    \end{split}
\end{align*}

If we have $M_a = M$, then the decryption process $D(E(M_a) \times E(M)^{-1} = D(1) = 0)$. The challenger (such as the insurance company) gets encrypted ciphertext from an off-chain IPFS storage place and decrypts the EHR data with a private key: $\lambda^{-1}~mod~n$. If the above result is 0, it indicates that the two plaintexts are the same so that the predictive claim could be settled. Otherwise, no claim can be determined. This operation allows the EHR data to be encrypted and out-sourced to off-chain IPFS storage for processing while keeping the privacy-preserving feature and resisting chosen-plaintext attacks from semi-trusted entities such as smart sensors and edge nodes.
\end{proof}

\subsection{Privacy Analysis}

As we mentioned before, the patient's EHR data is not directly saved into the blockchain. EHR data are encrypted and stored in IPFS utilizing the MA-ABE and Paillier homomorphic encryption mechanism. The blockchain transaction stores the access event and hash index for the EHR data. Even if the blockchain transaction is attacked, it will not reveal the patient's sensitive information since the ABSA scheme will safeguard the patient's privacy. Also, the inherent Merkle DAG data structure utilized in IPFS will increase the degree of data confusion, which again enhances the system security.


{Our proposed MA-ABE and ABSA mechanisms achieve fine-grained access control and privacy protection. Only data users who satisfy the ACL policy can decrypt patients' EHR records. The ABSA protects the private attribute information of both doctors and patients. The signature aggregation process will decrease both encryption and decryption time to improve the system's efficiency.}

{In the medical insurance claim process, the data user from the third party only receives the ciphertext of the patient's EHR insurance information, which again could not be decrypted under the construction of the Paillier homomorphic encryption scheme. The insurance company only gets the encrypted EHR ciphertext but cannot get the sensitive EHR information. However, based on Paillier's characteristics, they can still perform the additive/subtraction homomorphic operation, which safeguards patients' EHR information privacy during the data sharing and analyzing procedure.}

\section{Experiments}

This section describes the implemented EHR management system consisting of three primary modules: MA-ABE, ABSA,  and blockchain. The MA-ABE module was developed and tested on the OpenABE cryptography library.  The ABSA module was programmed utilizing the Hyperledger Ursa library.  The blockchain framework was developed with the Hyperledger Fabric blockchain. Both modules were deployed and experimented on a desktop with the 2.8 GHz Intel i5-8400 processor and 8GB of memory using the Ubuntu 18.04 operating system.

\subsection{MA-ABE Module}
\label{sec:cpabe_module}

 MA-ABE module performs the system setup, key generation, encryption algorithm, and decryption algorithm. These functionalities are programmed with the OpenABE~\footnote{https://github.com/zeutro/openabe}, a cryptographic library with attribute-based encryption implementations in C/C++ languages.
 
 For instance, Alice has attributes Female, Nurse, Floor=3, Respiratory-Specialist.
Charlie has attributes Male, Doctor, Floor=5, Cardiologist. Bob encrypts a
patient’s medical file for staff members matching the policy $(Doctor~OR ~Nurse)$
AND $(Floor >= 3~AND~Floor < 5)$. Alice c open this patient’s file but
Charlie cannot open the patient's file.

 \subsubsection*{Phase 1 -- System Setup}
Phase 1 initializes the system parameters with file name prefix ``org1" shown as below: 
\begin{verbatim}
    ./oabe_setup -s CP -p org1
\end{verbatim}
and the {\tt CP} stands for the cipher-policy.

 \subsubsection*{Phase 2 -- Key Generation}
 Phase 2 generates the key for both Alice and Charlie, the key generation process is shown below:
 \begin{verbatim}
  # Key Generation for Alice
     ./oabe_keygen -s CP -p org1 -i
     "Female|Nurse|Floor=3|RespSpecialist" 
     -o aliceCP
 \end{verbatim}
\begin{verbatim}
  # Key Generation for Charlie
     ./oabe_keygen -s CP -p org1 -i
     "Male|Doctor|Floor=5|Cardiologist" 
     -o charlieCP
\end{verbatim}
Note that in our proposed scheme, attributes are associated with different participants, and the access policies are associated with the ciphertext.

 \subsubsection*{Phase 3 -- Encryption}
 Phase 3 processes the encryption process and enforces the rule such that Alice can decrypt the ciphertext while Charlie cannot decrypt the ciphertext. 
 \begin{verbatim}
  # Floor range is (Floor > 2 and
  Floor < 5)
     ./oabe_enc -s CP -p org1 -e 
     "((Doctor or Nurse) and
     (Floor in (2-5)))" 
     -i input.txt -o input.cpabe
 \end{verbatim}
 \subsubsection*{Phase 4 -- Decryption}
 Phase 4 process the decryption process for both Alice and Charile shown as below:
 \begin{verbatim}
  # Decrypt using Charlie's key
     ./oabe_dec -s CP -p org1 -k 
     charlieCP.key -i input.cpabe 
     -o plainFail.input.txt
 \end{verbatim}
 \begin{verbatim}
  # Decrypt using Alice's key 
     ./oabe_dec -s CP -p org1 -k
     aliceCP.key -i input.cpabe 
     -o plainOK.input.txt
 \end{verbatim}
 Based on the pre-defined ACL policy, the decryption process with Charlie's key will fail, while the decryption process with Alice's key will succeed.

\subsection{ABSA Module}
\label{ABSA module}

As shown in Fig. \ref{fig:workflow}, ABSA performs key generation processes, signing, signature and verification key aggregation, and verification of aggregated signatures. These functionalities are programmed with the Hyperledger Ursa \cite{hyperledgerursa}, a cryptographic library for developing blockchain-based applications. 
The ABSA module operates in the following six steps, as shown in Fig. \ref{fig:ABSA-process}. 


\subsubsection*{Phase 1 -- Initialization}
\begin{figure}[t]
\centering
\includegraphics[width=0.46\textwidth]{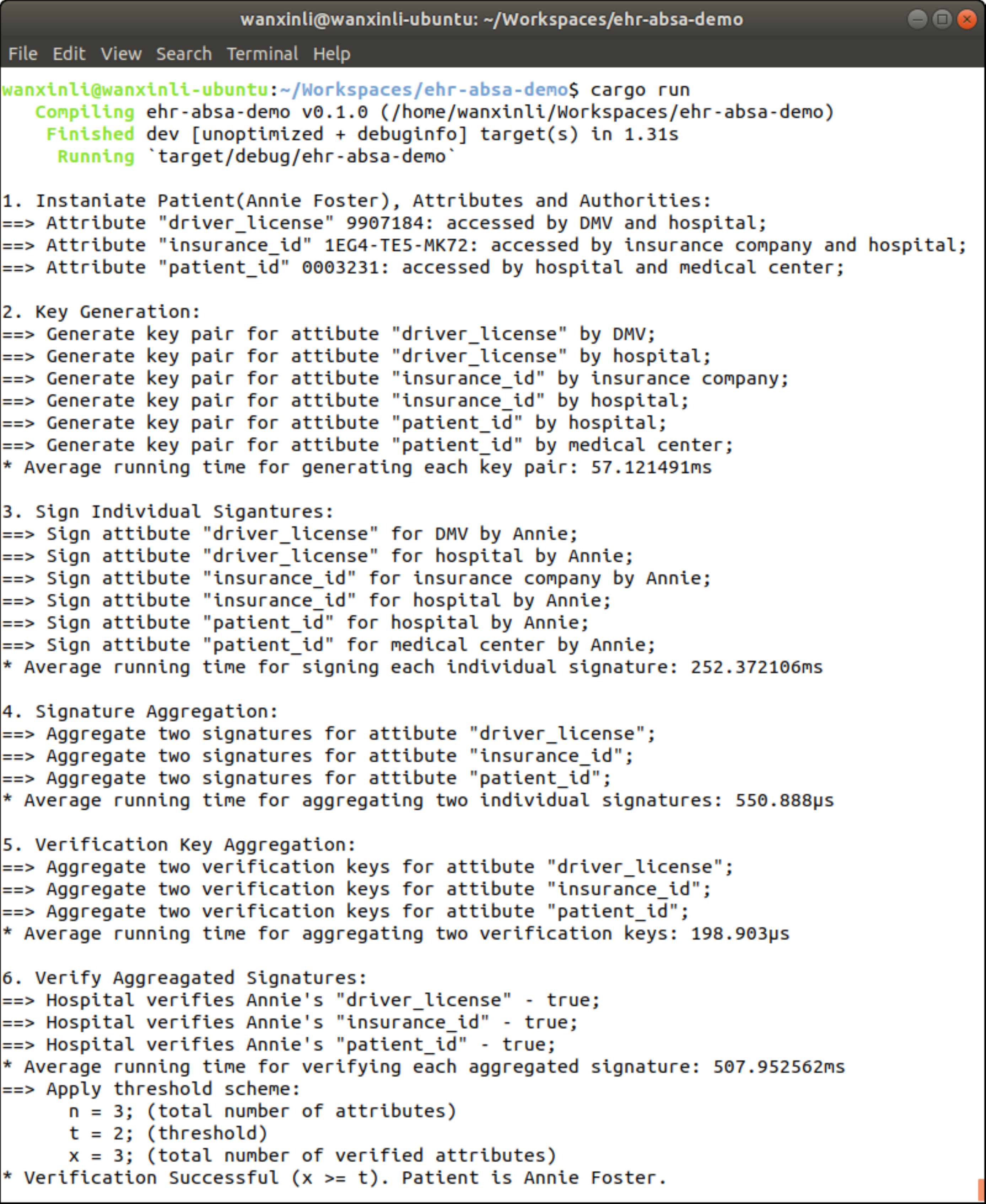}
\caption{Process of the ABSA module.}
\label{fig:ABSA-process}
\end{figure}

Phase 1 initializes the participant instances, including a patient with multiple attributes and multiple authorities. For example, as shown in Fig. \ref{fig:ABSA-process}, patient Annie Foster obtain three attributes, which are {\tt driver\_license} (value: {\tt 9907184}) that can be accessed by Department of Motor Vehicles (DMV) and hospital, {\tt insurance\_id} (value: {\tt 1EG4-TE5-MK72}) that can be accessed by her insurance company and hospital, and {\tt patient\_id} (value: {\tt 0003231}) that can be accessed by the hospital and medical center. 

\subsubsection*{Phase 2 -- Key Generation}
In this step, every attribute authority generates the key pair for attributes managed by itself. We use the BLS signature scheme \cite{boneh2001short} to construct the key pair generator, which generates the verification key for the data user and the signing key for the data owner (patient), as shown below:


\begin{verbatim}
let generator = normal::Generator:
                :generator();
let (verify_key, sign_key) = normal:
                :generate(&generator);
\end{verbatim}

\subsubsection*{Phase 3 -- Signing Individual Signatures}
In Phase 3, the data owner applies the signing keys to sign hashed attribute values and saves aggregated signatures in the data owner's private profile. By utilizing the BLS signature mechanism \cite{boneh2001short}, every signature contains two elements on the elliptic curve. The average running time for this signature signing process is about 252 ms.


\subsubsection*{Phase 4 -- Signature Aggregation}
For each attribute, the module automatically aggregates multiple individual signatures from Phase 3 into one signature, 
which is saved on the blockchain in a hexadecimal format of two points from an elliptic curve. The result indicates that the average execution time for aggregating two signatures is about 0.5 ms.


\subsubsection*{Phase 5 -- Verification Key Aggregation}
For each attribute, the module automatically aggregates multiple verification keys from Phase 2 into one short verification key and saves it on the blockchain. The result indicates that the average running time for aggregating two verification keys is around 0.2 ms.

\subsubsection*{Phase 6 -- Verify Aggregated Signatures}
\label{sec:phase6}

Finally, the data user can verify every aggregated signature saved on a blockchain transaction. The verification function takes the hashed attribute value, an aggregated verification key, and an associated generator as inputs for the aggregated signature and uses BLS pairing \cite{boneh2003survey} to validate  aggregated signatures:


\begin{verbatim}
let result = aggregated_signature.verify(
             h(attribute_value), None, 
             &aggregated_verification_key, 
             &generator);
\end{verbatim}

The average running time for verifying every aggregated signature is about 508 ms. Next, data users could verify  data owners (patients) by applying a multi-signature threshold scheme, and the total numbers of the aggregated signatures will be compared with the pre-defined threshold.


\subsection{Blockchain Module}
\label{sec:blockchain_module}

We develop and deploy blockchain modules on the Hyperledger Fabric platform. Patients' GIDs, first and the last names, generated ABSA signatures, and the one-time self-destructing url addresses are saved in a private storage place that can only be accessed by doctors who satisfy access control policies. We utilize {\tt https://1ty.me/}~\footnote{https://1ty.me/} to encode EHR data addresses saved in edge storage. Once the {\tt 1ty.me} address has been reviewed,  it will become invalid and cannot be reaccessed to retrieve EHR information.


\begin{figure}[t]
\centering
\includegraphics[width=0.43\textwidth]{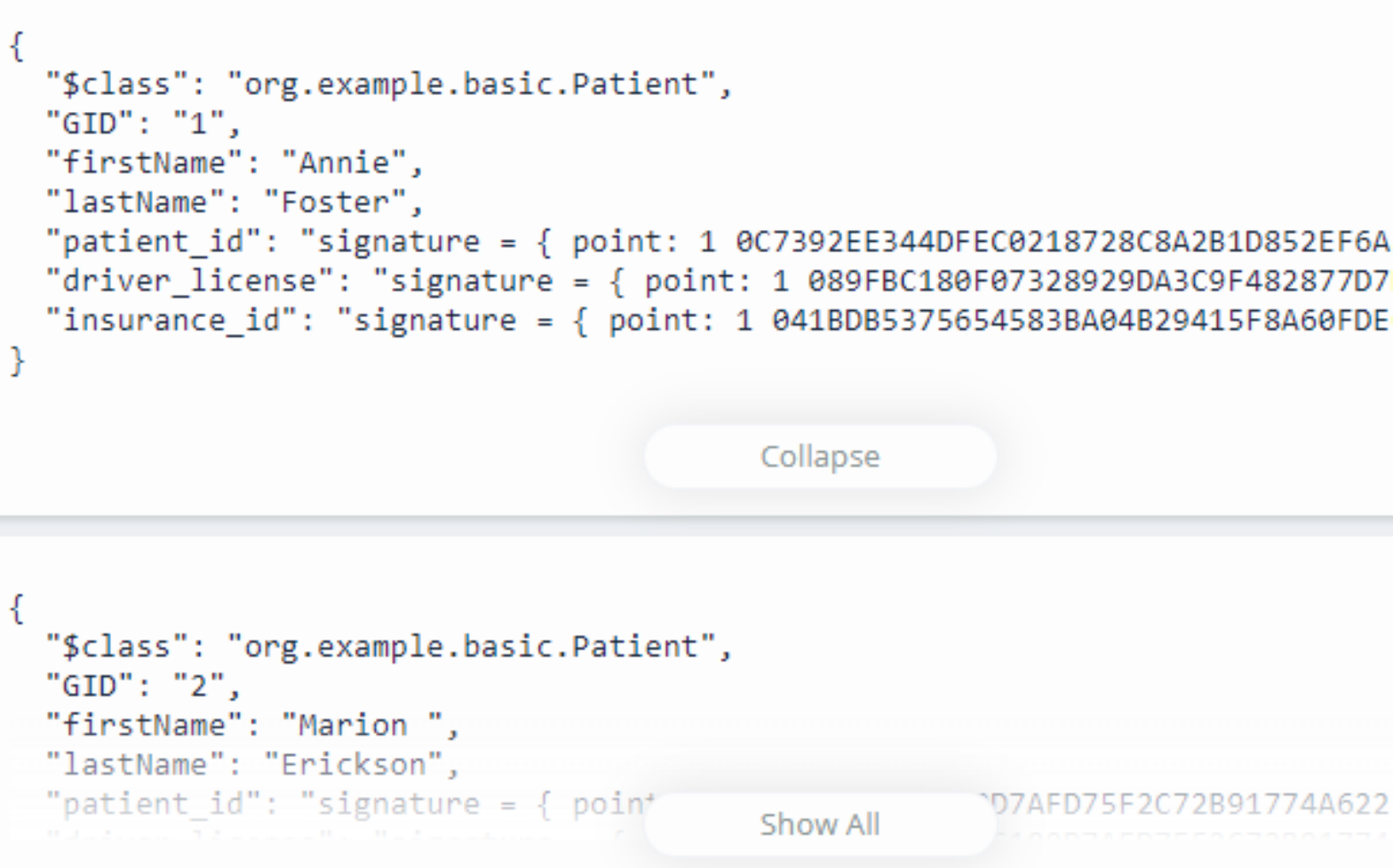}
\caption{A doctor can access a list of patient profiles.}
\label{fig:doctor-access}
\end{figure}

\begin{figure}[t]
\centering
\includegraphics[width=0.43\textwidth]{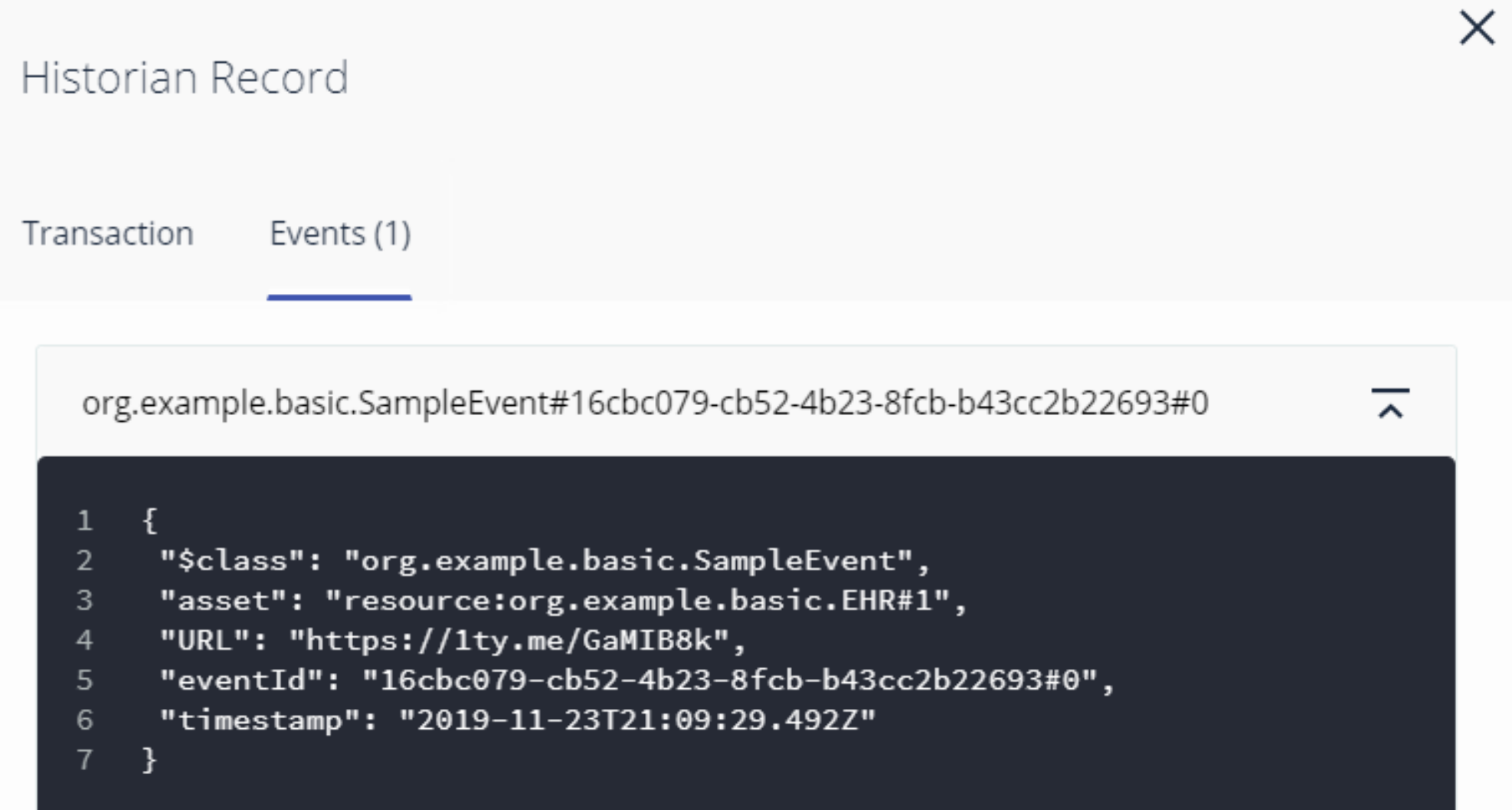}
\caption{Result of executing the smart contract.}
\label{fig:transaction-detail}
\end{figure}


\subsubsection{Test of Access Control on Doctor}

We conduct the following experiments to evaluate the ACL mechanism. As shown in Fig. \ref{fig:doctor-access}, a doctor can access patients' information, including GIDs, first names, last names and generated ABSA signatures by satisfying access policies defined in ACL policy. When the doctor would like to retrieve the patient’s EHR information from edge nodes, they first need to validate the patient's identity by checking the set of aggregated signatures against the pre-defined threshold. Next, the system will return the one-time {\it url} of  EHR information saved in edge storage. Meanwhile, the blockchain system permanently saves the access event as the new blockchain transaction, including event ID, timestamp, and other information, as shown in Fig. \ref{fig:transaction-detail}. Therefore,  people can check all existing EHR access events for further investigation and data provenance.

\subsubsection{Test of Access Control on Patient}



In this evaluation, we switch participants' identities to the data owner who could access their EHR information. Unlike the role of the doctor, a patient cannot access another patient's sensitive information. For example, if patient Carmen Maxwell ({\em GID}: 3) tries to retrieve the EHR data of patient Laverne Green ({\em GID}: 4), our framework will automatically deny this transaction request. 



\section{Performance Evaluation and Comparison}

\subsection{MA-ABE Execution Cost}

To evaluate the multi-authority CP-ABE module's performance, we discussed the MA-ABE detailed construction phases in Section \ref{sec:cpabe_module}. In addition, we conducted extensive experiments to analyze the effect of changing access policies for MA-ABE's key generation, encryption, and decryption process. 



 We change the access policies for each user (e.g., Alice and charlie) to obtain both encryption and decryption time. For instance, the access policy is defined as: $(Doctor~or~Nurse)$ and $(Floor~in~(2-5))$ as boolean formulas. We conduct three rounds of experiments with different access policies and found that the access policy significantly impacts both encryption and decryption time. 

\begin{figure}[t]
\centering
\includegraphics[width=0.34\textwidth]{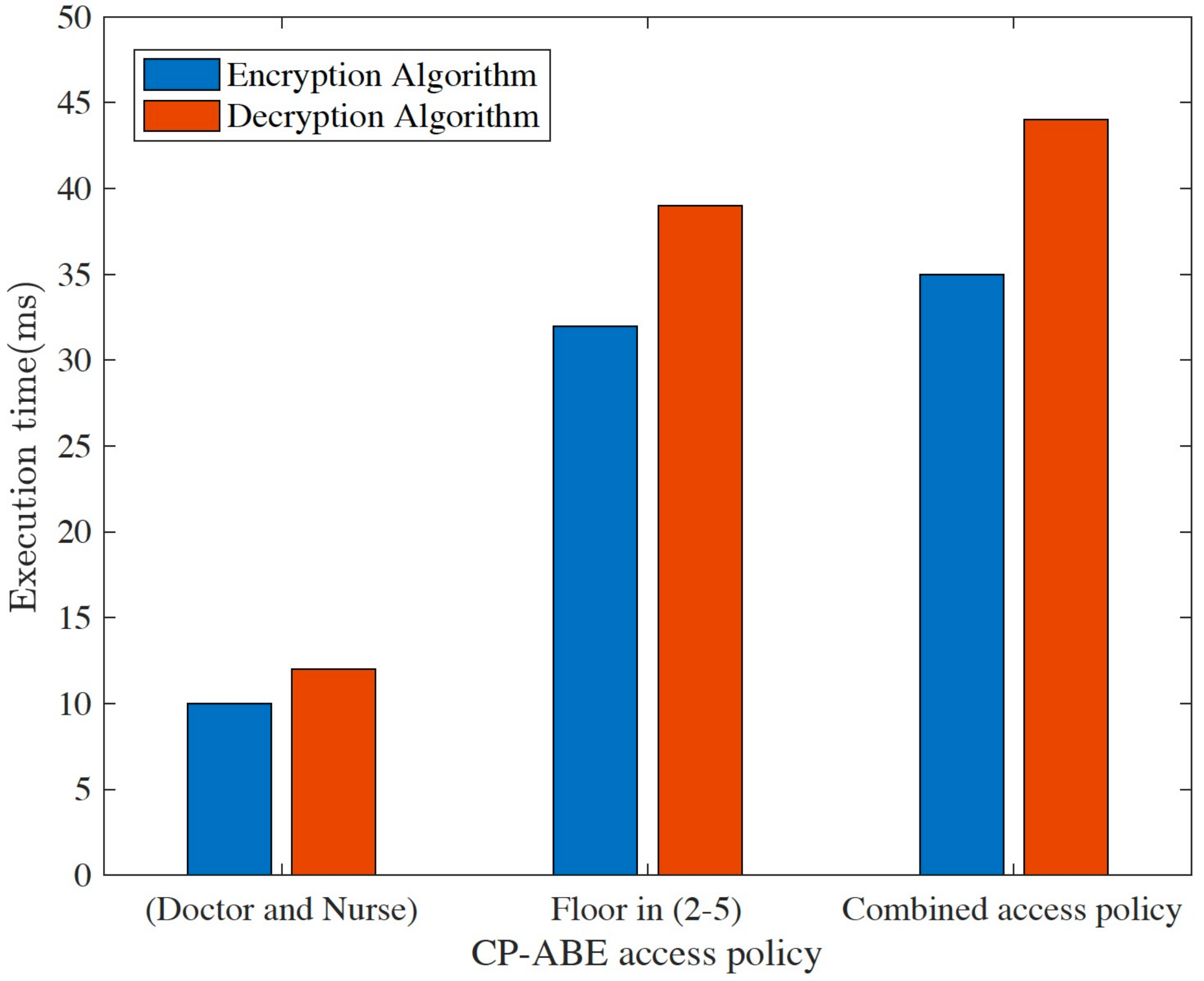}
\caption{Execution time of MA-ABE encryption and decryption algorithms vs. different access policies.}
\label{fig:cpabe2}
\end{figure}

As shown in Fig.~\ref{fig:cpabe2}, encryption and decryption time of access policy $(Doctor~or~Nurse)$ is 10ms and 12ms, the encryption and decryption time of $(Floor~in~(2-5))$ is 32ms and 39ms, and  the encryption and decryption time of combined access policy  $(Doctor~or~Nurse)$ {\it and} $(Floor~in~(2-5))$ is 34ms and 43ms. We argue that the simple boolean formulas such as the $(Doctor~or~Nurse)$ will take the least execution time when compared to the more complicated boolean formulas such as $(Floor~in~(2-5))$, and the result from the combined access policy $(Doctor~or~Nurse)$ {\it and} $(Floor~in~(2-5))$ indicates that the encryption and decryption execution time heavily depends on latter access policy $(Floor~in~(2-5))$.

\subsection{Homomorphic Encryption Execution Cost}

\begin{figure}[t]
\centering
\includegraphics[width=0.34\textwidth]{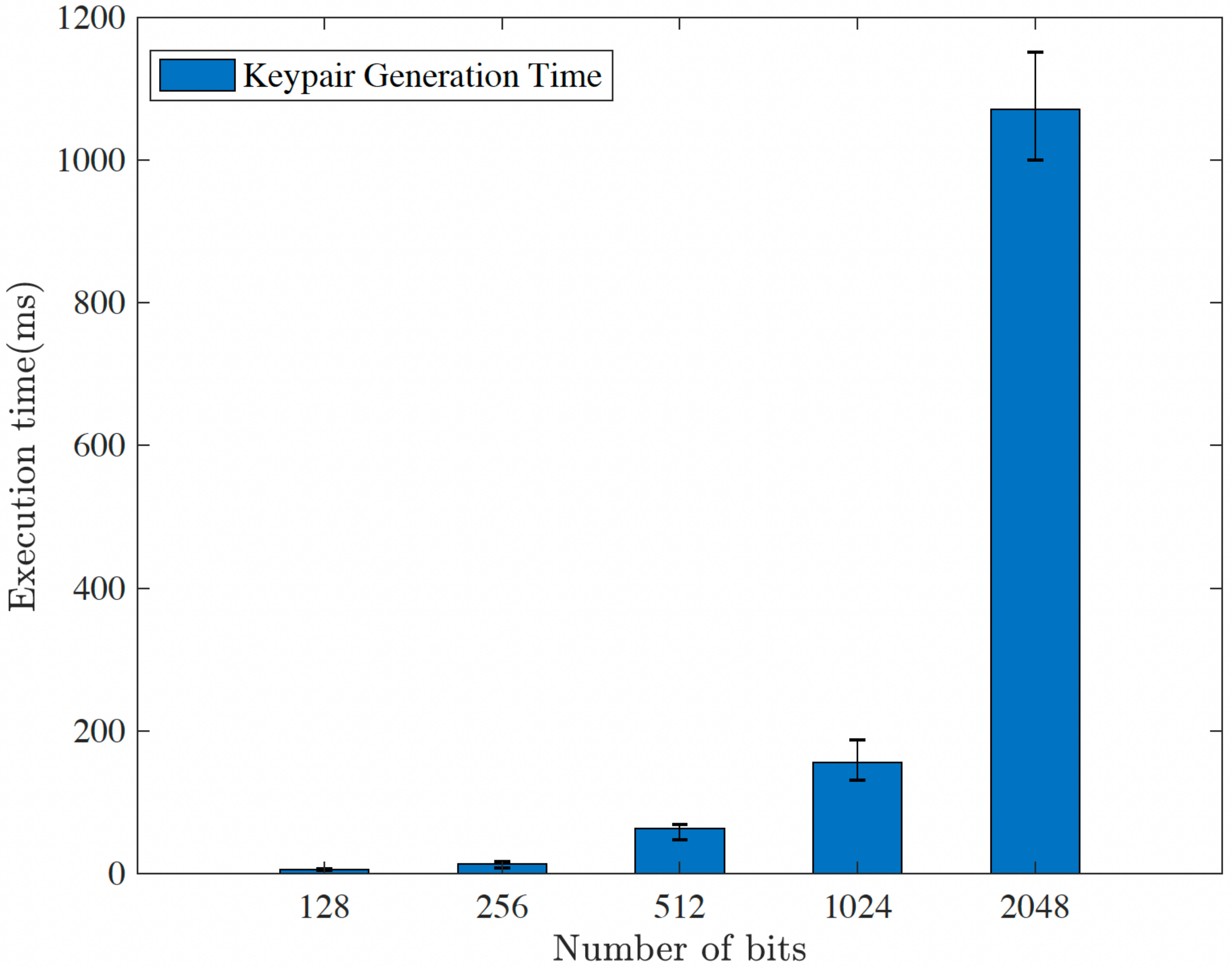}
\caption{Execution time of HE key generation vs. number of bits.}
\label{fig:hekey}
\end{figure}

{To evaluate the Paillier cryptosystem's performance, we conducted the experiments based on the jspaillier library, a Javascript implementation of the Paillier homomorphic encryption scheme. We increased the number of bits from 128 to 256, 512, 1024, and 2048, and measured the execution time of key pair generation processes. As we can observe from Fig.~\ref{fig:hekey}, the key generation time is 5 ms for 128 bits, 13 ms for 256 bits, 34 ms for 512 bits, 161 ms for 1024 bits, and 1071 bits for 2048 ms. It indicates that the key generation time will grow exponentially when the number of bits grows. Paillier takes longer when generating the key pair than other PKI-based encryption schemes. We should keep the generator relatively small in a real-world scenario to reduce the key generation time.}

\begin{figure}[t]
\centering
\includegraphics[width=0.34\textwidth]{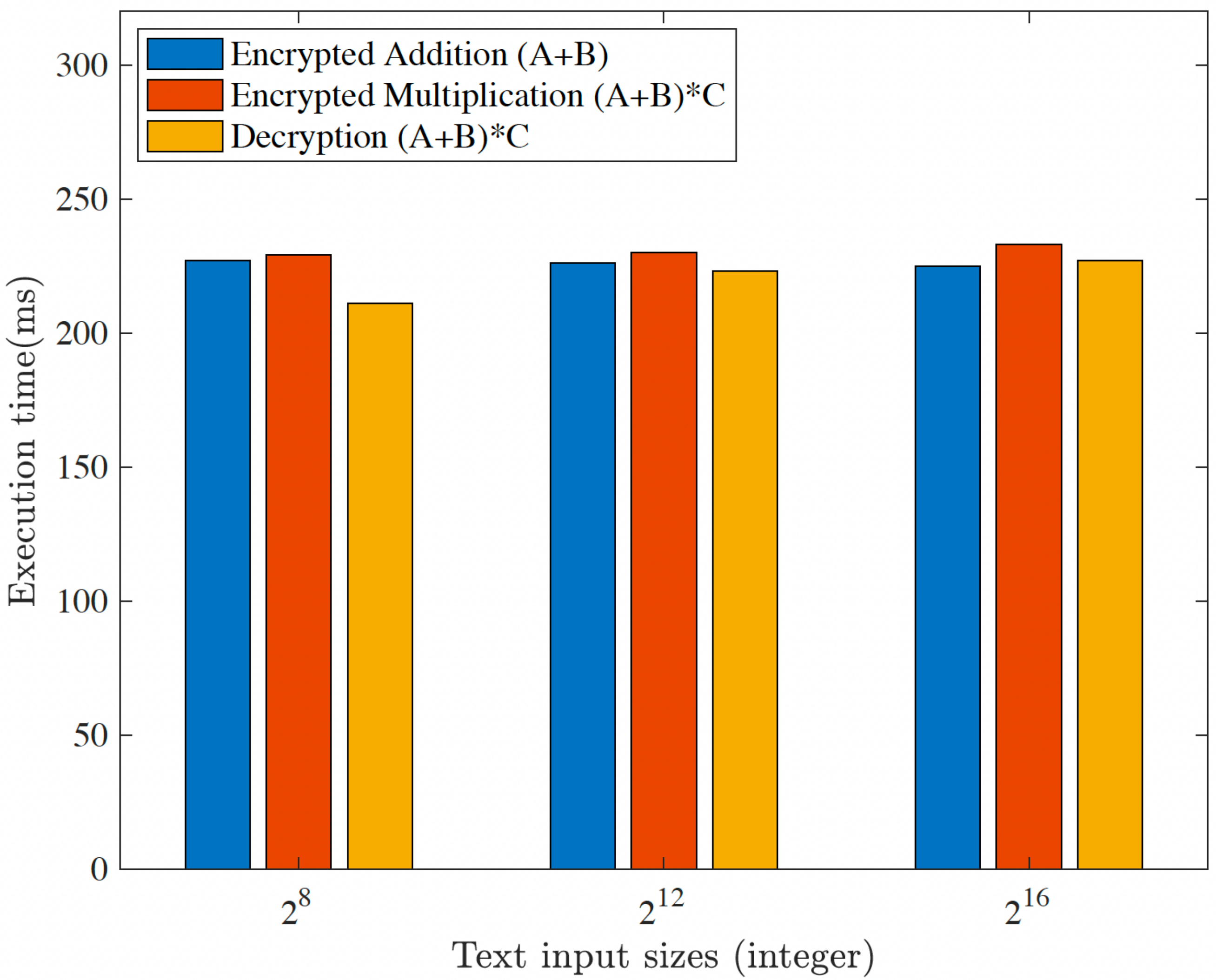}
\caption{Execution time of Paillier cryptosystem encrypted addition, multiplication, and decryption algorithm.}
\label{fig:heencdec}
\end{figure}

Next, we measure the execution time of the Paillier encrypted addition (formula is $(A+B)$), encrypted multiplication ($(A+B)*C$), and decryption ($(A+B)*C$) processes. As shown in Fig.~\ref{fig:heencdec}, all encryption and decryption phases have relatively constant time (within milliseconds) when changing the text input sizes. For instance, the encrypted addition and encrypted multiplication time for $2^{12}$ is 226 ms and 230 ms, and the decryption time is 223 ms. It shows that the input text sizes will not affect encryption and decryption time for the Paillier encryption scheme. Therefore, our proposed system can achieve high performance when increasing the input text size.

\subsection{ABSA Aggregation Execution Cost}

\begin{figure}[t]
\centering
\includegraphics[width=0.34\textwidth]{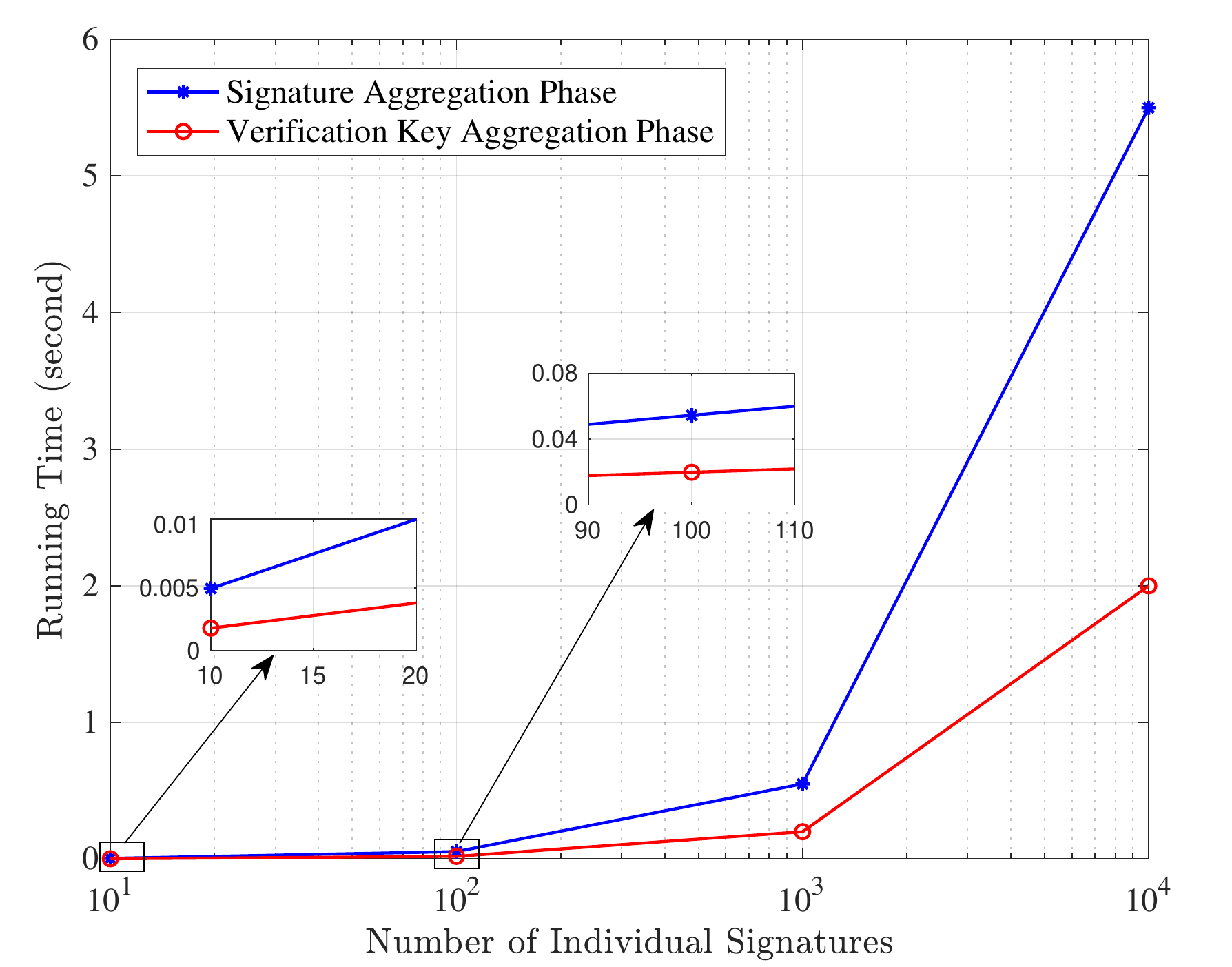}
\caption{Running time of aggregation phases vs. the number of individual signatures.}
\label{fig:aggregation-time}
\end{figure}

To evaluate the ABSA module's performance, we have discussed the average running time for each phase in Section \ref{ABSA module}. In addition, we performed extensional experiments to analyze the effect of changing the number of individual signatures. We increased the number of individual signatures from 10 to 100, 1000, and 10,000, and measured the overall running time of signature aggregation, verification key aggregation, and final verification phases. Both the running time of signature aggregation and verification key aggregation processes had linear growth with growing individual signatures (Fig. \ref{fig:aggregation-time}). Note that the x-axis in the figure is in a logarithmic scale. Compared to the other phases, the aggregation phases for signatures and verification keys take less time in milliseconds. Therefore, our ABSA module can preserve high overall performance while increasing the number of individual signatures.

\begin{figure}[t]
\includegraphics[width=0.35\textwidth]{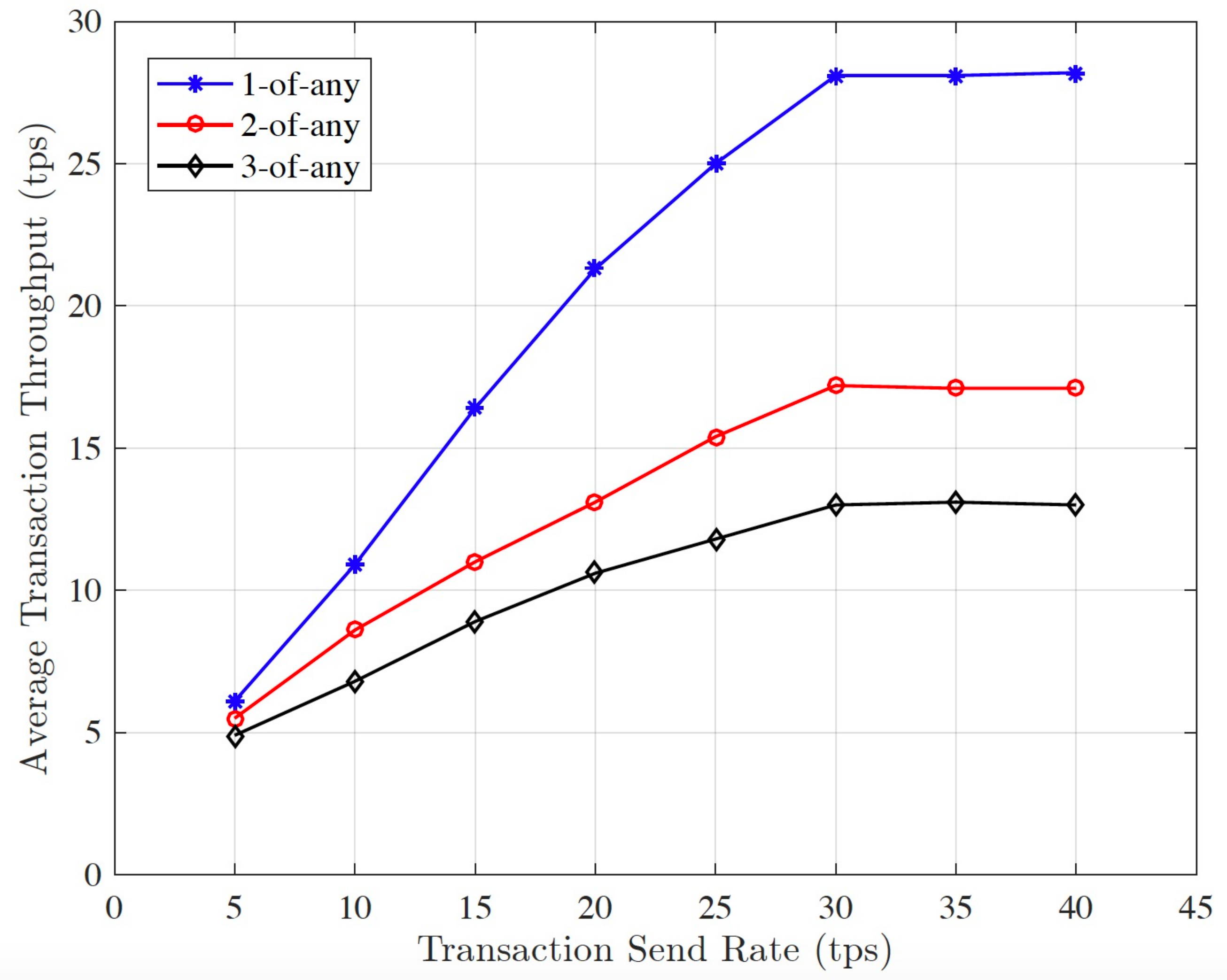}
\centering
\caption{Transaction throughput with different endorsement policies when changing the transaction send rate.}
\label{fig:tsrtps}
\end{figure}

\begin{figure}[t]
\includegraphics[width=0.35\textwidth]{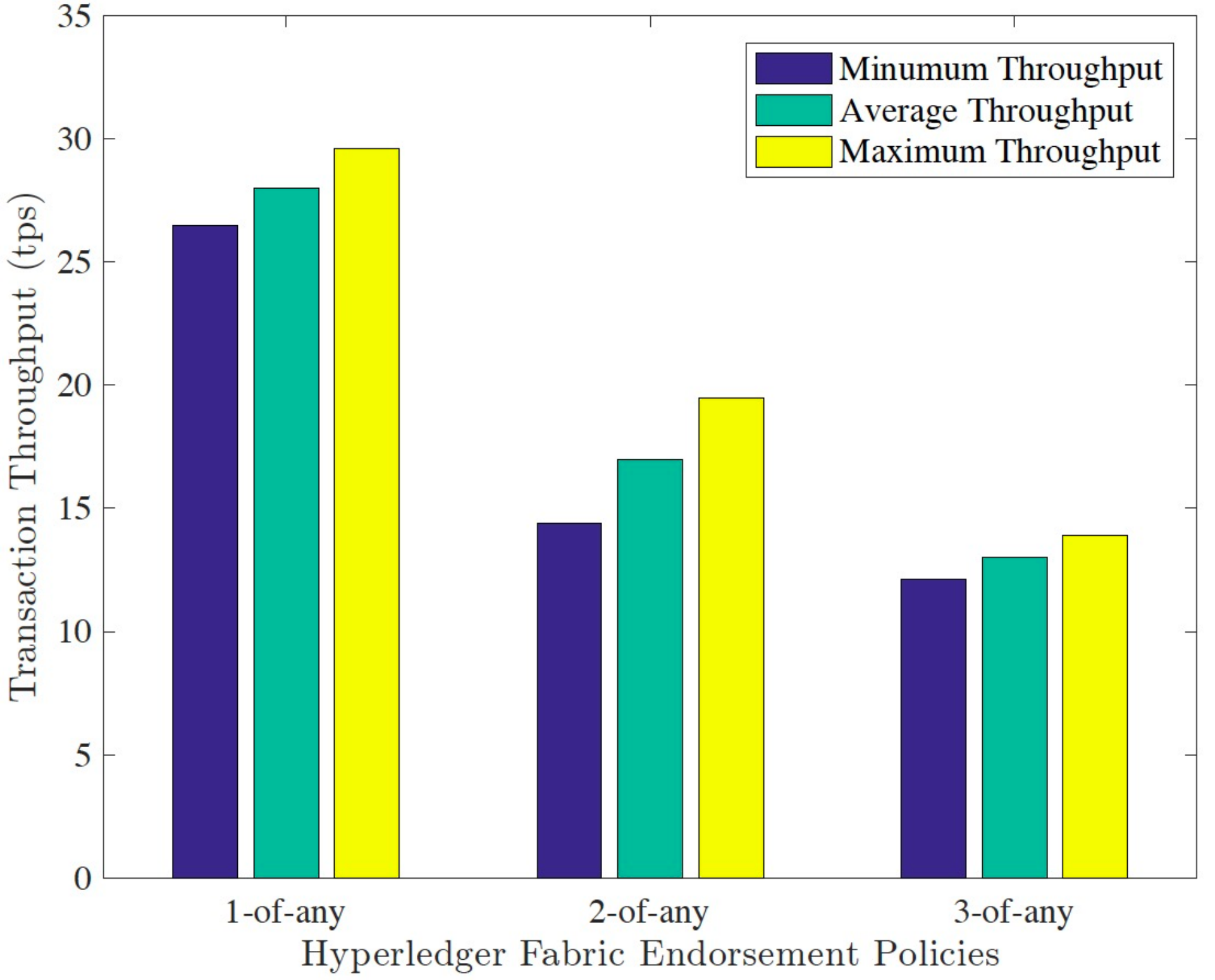}
\centering
\caption{Minimum, average and maximum transaction throughput vs. Hyperledger Fabric endorsement policy under the send rate of 30 tps.}
\label{fig:tps}
\end{figure}

\begin{figure}[t]
\includegraphics[width=0.35\textwidth]{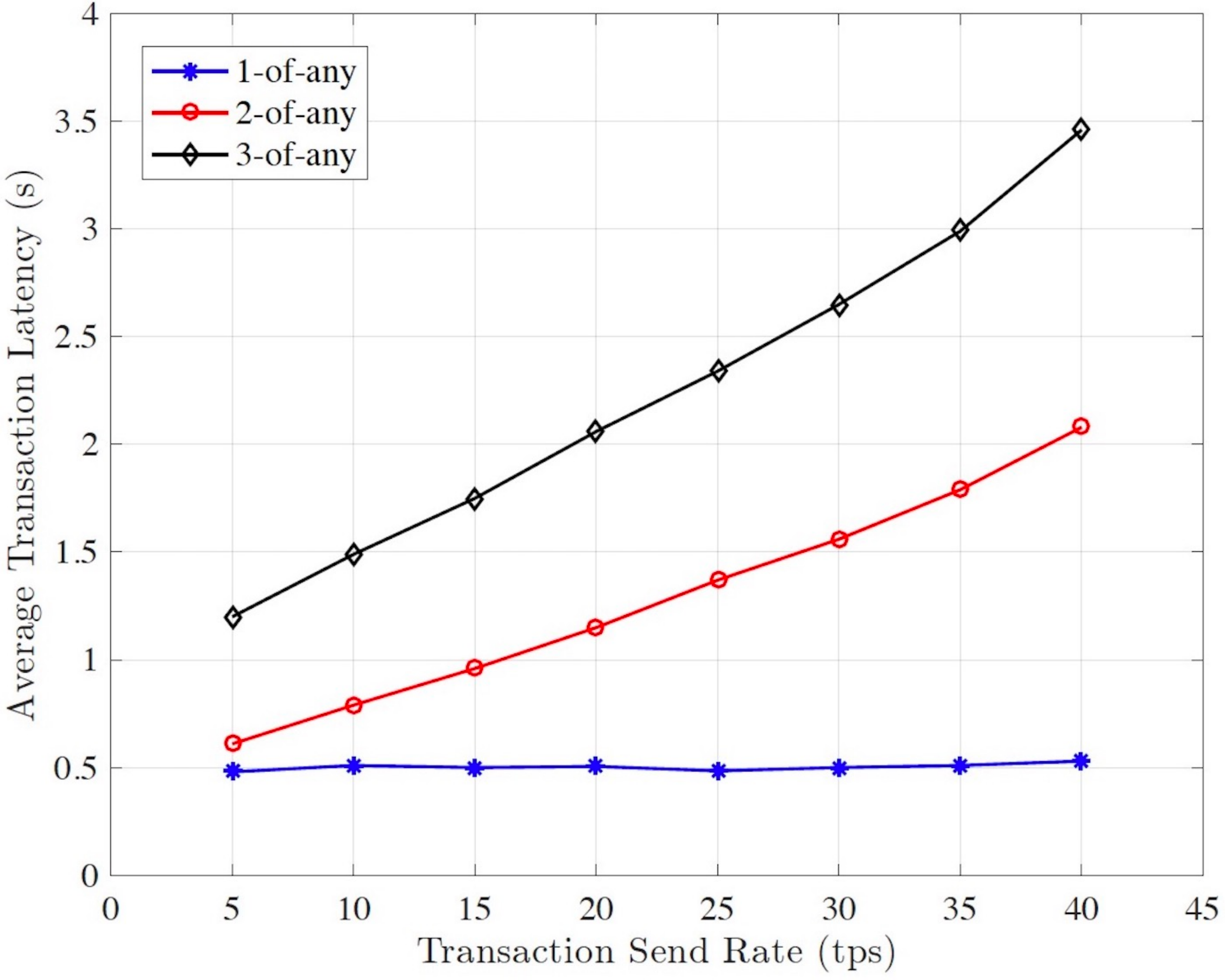}
\centering
\caption{ Transaction latency with varying endorsement policy when changing transaction send rate.}
\label{fig:tsrlatency}
\end{figure}

\begin{figure}[t]
\includegraphics[width=0.35\textwidth]{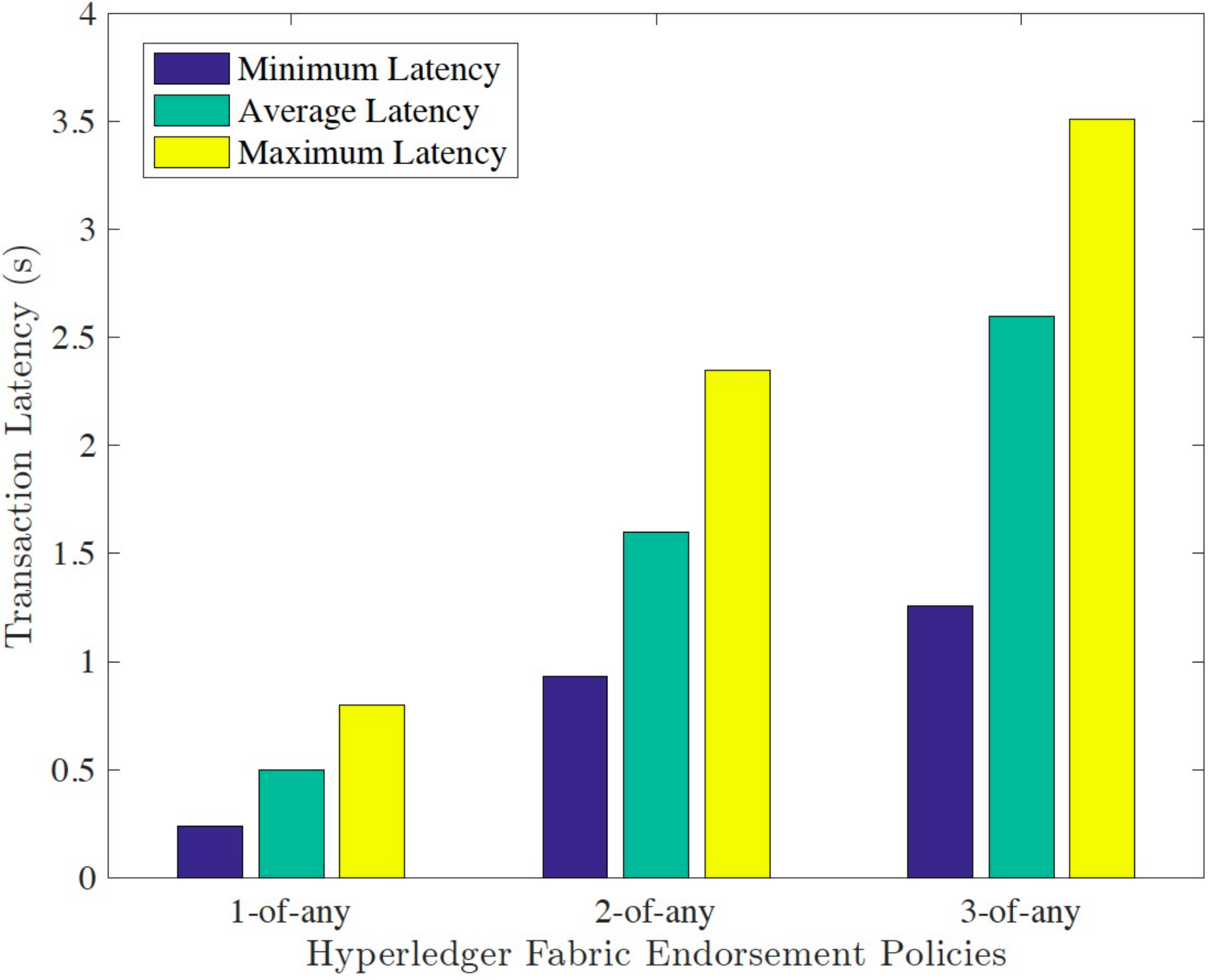}
\centering
\caption{Minimum, average and maximum transaction latency vs. Hyperledger Fabric endorsement policy under the send rate of 30 tps.}
\label{fig:latency}
\end{figure}

\subsection{Transaction Throughput}

Transaction throughput indicates how many transactions can be verified and processed per second. We measure blockchain batch transaction throughput results under multiple endorsement policies and transaction send rates as shown in Fig.~\ref{fig:tsrtps} and Fig.~\ref{fig:tps}. In Fig.~\ref{fig:tsrtps}, when we increase the send rate of blockchain batch transactions, the average transaction throughput rises dramatically in the initial phase and later becomes stable at 27 tps, 17 tps, and 13 tps under the 1-of-any, 2-of-any, and 3-of-any endorsement policies after 30tps transaction send rates. In Fig.~\ref{fig:tps}, our blockchain network has 27.5 tps, 17.4 tps, and 13 tps under 1-of-any, 2-of-any, and 3-of-any endorsement policy when transaction send rates have the fixed-rate 30tps, respectively. As observed from the experiment, the average transaction throughput decreases dramatically when the blockchain system increases the number of peers participating in the endorsement procedure. The reason is that more endorsing peers increase the complexity and overhead of the transaction endorsement procedure.


\subsection{Transaction Latency}

{Transaction latency measures the time for an issued transaction from being submitted to processed on the ledger. All experiments are conducted based on different endorsement policies and transaction send rates: 1-of-any, 2-of-any, and 3-of-any. In Fig.~\ref{fig:tsrlatency} when we increase the transaction send rate, average transaction latency also increases dramatically for 2-of-any and 3-of-any endorsement policies. However, average transaction latency remains steady at around 0.5 s under the 1-of-any endorsement scheme when the transaction sends rate is below 30 tps. After passing 30 tps transaction sends rates, the latency for the 1-of-any endorsement policy grows slowly. The reason is that the 1-of-any endorsement policy has a  relatively high threshold of transaction throughput  because of fewer peers. As we can see from Fig.~\ref{fig:latency}, we conduct the experiments under 30tps send rates with an increasing number of endorsing peers from 1 to 6; both minimum, maximum,  and the average latency results all grow dramatically.}

\begin{table*}[h]
\centering
\caption{Comparisons with other blockchain-based EHR systems.}
\label{tab: consensus comparison}
\begin{tabular}{ccccccc}
Proposed Scheme & \thead{Blockchain Type} & \thead{Access Control} & \thead{Data Encryption} & \thead{Off-chain Storage} & \thead{Consensus Protocol} & Blockchain Platform \\ \hline
EHRChain \cite{li2021ehrchain}  & Permissioned       & SHDPCPC-ABE  & CP-ABE & IPFS & PBFT & Hyperledger Fabric \\ \hline
Healthchain \cite{xu2019healthchain}   & Permissionless       & $\times$  & AES & $\times$ & PBFT & Bitcoin Like \\ \hline
Fortifiedchain \cite{egala2021fortified}  & Permissioned & Ring-based AC  & $\times$ & Edge+IPFS & PoS & Ethereum \\ \hline
Medge-chain \cite{abdellatif2021medge}  & Permissioned &  $\times$ & $\times$ & Edge &  $\times$ &  $\times$ \\ \hline
BEdgeHealth \cite{nguyen2021bedgehealth}  & Permissioned &  $\times$ & $\times$ & Edge &  $\times$ &  $\times$ \\ \hline
BIoTHR \cite{ray2021biothr}  & $\times$ & Swarm Exchange  & PKI & Swarm Node & $\times$ & $\times$ \\ \hline
ABMS-EHR \cite{guo2020icbc}  & Permissioned & ABMS  & MA-ABE & Edge & Kafka & Hyperledger Fabric \\ \hline
Our Scheme  & Permissioned & MA-ABSA  & MA-ABE+HE  & Edge+IPFS & Kafka & Hyperledger Fabric \\ \hline
\end{tabular}
\end{table*}

\subsection{Comparison with Other Blockchain-based EHR Systems}
In this subsection, we compare the different blockchain types, access control methods, data encryption algorithms, off-chain storage architecture, consensus protocol, and the blockchain platform among our proposed scheme and other state-of-the-art blockchain-based EHR data management systems including the EHRChain \cite{li2021ehrchain}, Healthchain \cite{xu2019healthchain}, Fortifiedchain \cite{egala2021fortified}, Medge-chain \cite{abdellatif2021medge}, BEdgeHealth \cite{nguyen2021bedgehealth}, BIoTHR \cite{ray2021biothr}, and ABMS-EHR \cite{guo2020icbc}. 

As we observe from Table \ref{tab: consensus comparison}, most of the proposed blockchain-based EHR systems utilize the permissioned blockchain type, and the edge node architecture is popular among the off-chain storage. Only EHRChain~\cite{li2021ehrchain}, ABMS-EHR~\cite{guo2020icbc}, and our proposed system offer cryptographic features, and our ABSA scheme has the public key aggregation and signature threshold features. The integrated MA-ABE with homomorphic encryption can protect data security in a more efficient and end-to-end privacy-preserving way. We provide detailed comparative experiments with EHRChain~\cite{li2021ehrchain} and ABMS-EHR~\cite{guo2020icbc} in the performance evaluation section. For the Fortifiedchain~\cite{egala2021fortified}, Medge-chain \cite{abdellatif2021medge}, BEdgeHealth \cite{nguyen2021bedgehealth}, and BIoTHR \cite{ray2021biothr}, we conduct the table comparison results since their evaluation is based on IPFS file download and upload operations with different file sizes and other evaluation metrics.

\begin{figure}[t]
\centering
\includegraphics[width=0.36\textwidth]{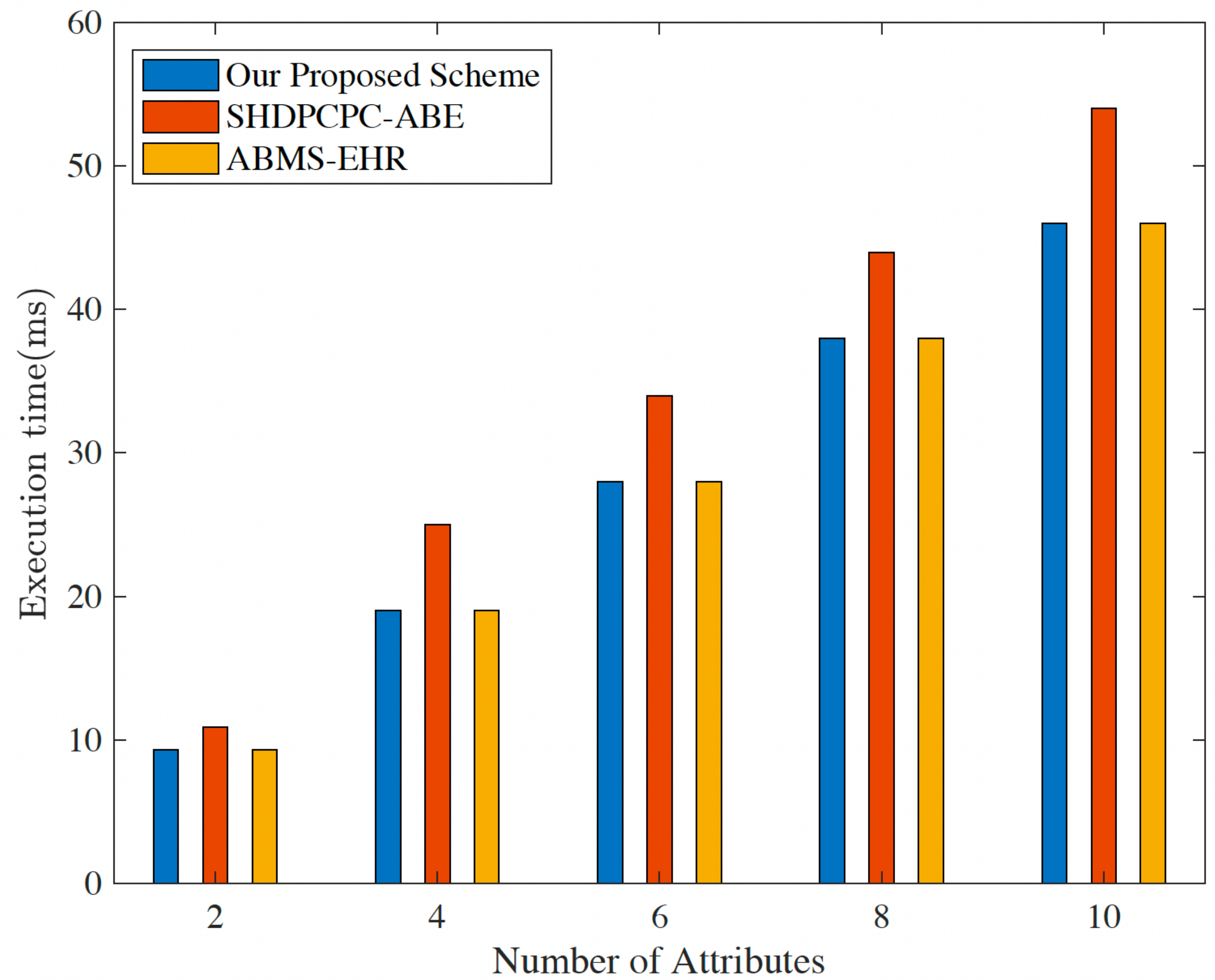}
\caption{Execution time of MA-ABE key generation of our scheme, SHDPCPC-ABE~\cite{li2021ehrchain}, and ABMS-EHR~\cite{guo2020icbc}.}
\label{fig:abekeygen}
\end{figure}

In the first experiment, we change the number of attributes for each user to measure the key generation process time for our scheme, EHRChain~\cite{li2021ehrchain}, and ABMS-EHR~\cite{guo2020icbc}.
We increased the number of attributes from 2 to 4, 6, 8, and 10. For instance, the attribute value can be:  $\it{Gender (Male/Female)}$, $\it{Job Title (Doctor/Nurse)}$, $\it{Location (Floor=3/ Floor= 5)}$, and $\it{Category(Respiratory specialist/ Cardiologist})$. As we can observe from Fig.~\ref{fig:abekeygen}, the execution time of the key generation process grows linearly with the increasing number of attributes. When the attributes number is 10, the process of key generation execution time is roughly 47ms for our scheme and ABMS-EHR~\cite{guo2020icbc}, and 55ms for EHRChain~\cite{li2021ehrchain}. The results indicate that our scheme is feasible and efficient when generating the attribute keys.
The blue color represents the execution time of our scheme, and the red and yellow color shows the SHDPCPC-ABE~\cite{li2021ehrchain} and ABMS-EHR~\cite{guo2020icbc} scheme. 

\begin{figure}[t]
\centering
\includegraphics[width=0.36\textwidth]{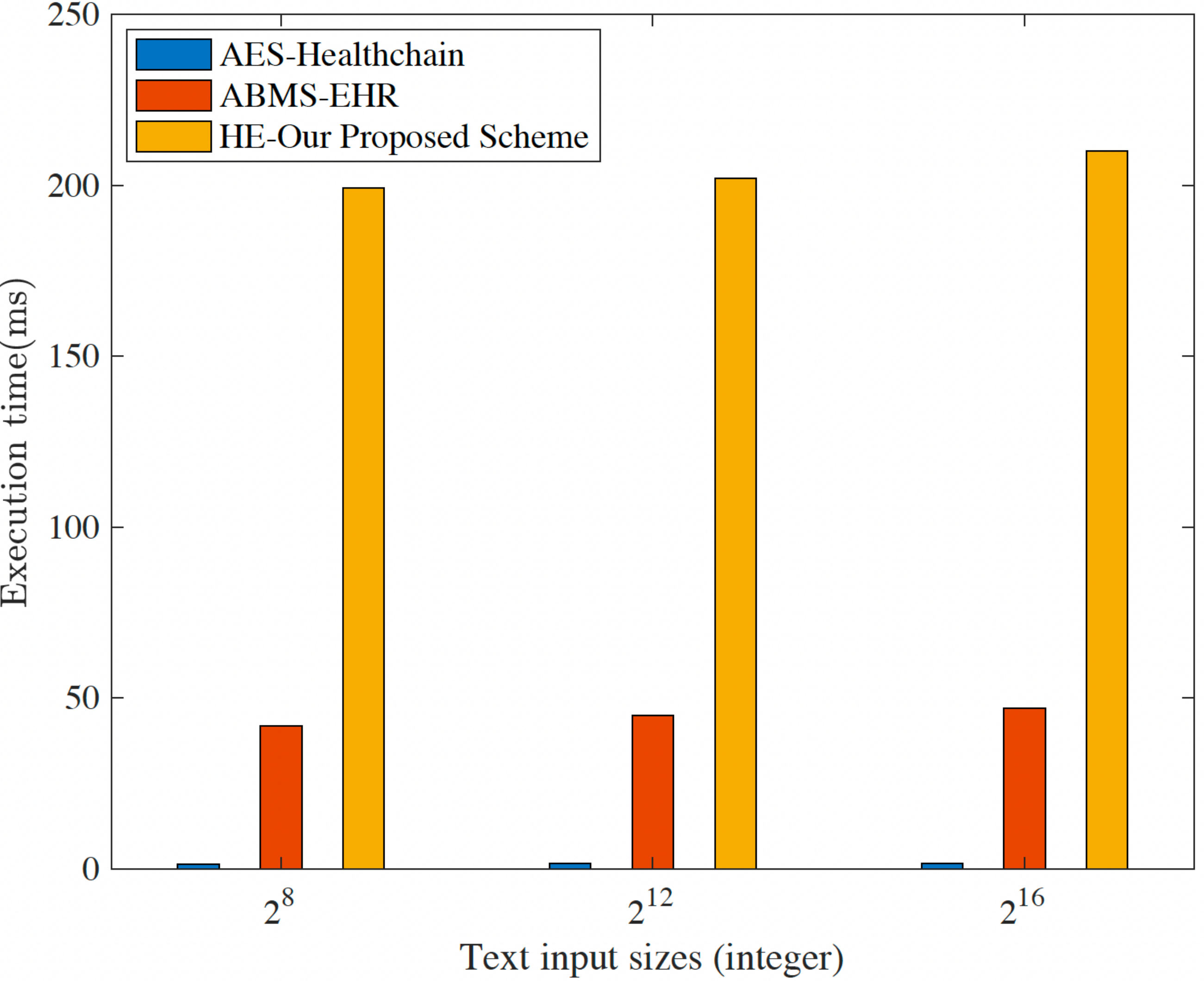}
\caption{Execution time of HE encryption of our scheme, AES-Healthchain~\cite{xu2019healthchain}, and ABMS-EHR~\cite{guo2020icbc}.}
\label{fig:heabeaes}
\end{figure}

Next, we measure the running time of the HE encryption operation and compare it with the AES-Healthchain~\cite{xu2019healthchain} and ABMS-EHR~\cite{guo2020icbc}. As shown in Fig. \ref{fig:heabeaes}, the encryption time of the all mentioned scheme will remain constant when increasing the input text sizes. For instance, AES, ABE, and HE schemes roughly take 1.5ms, 45ms, and 203ms for the encryption process. Accordingly, our proposed method utilized the Paillier homomorphic encryption scheme to provide end-to-end privacy. The execution time exceeds the AES-Healthchain and ABMS-EHR but still satisfies the real-world response requirement.

\begin{figure}[t]
\centering
\includegraphics[width=0.36\textwidth]{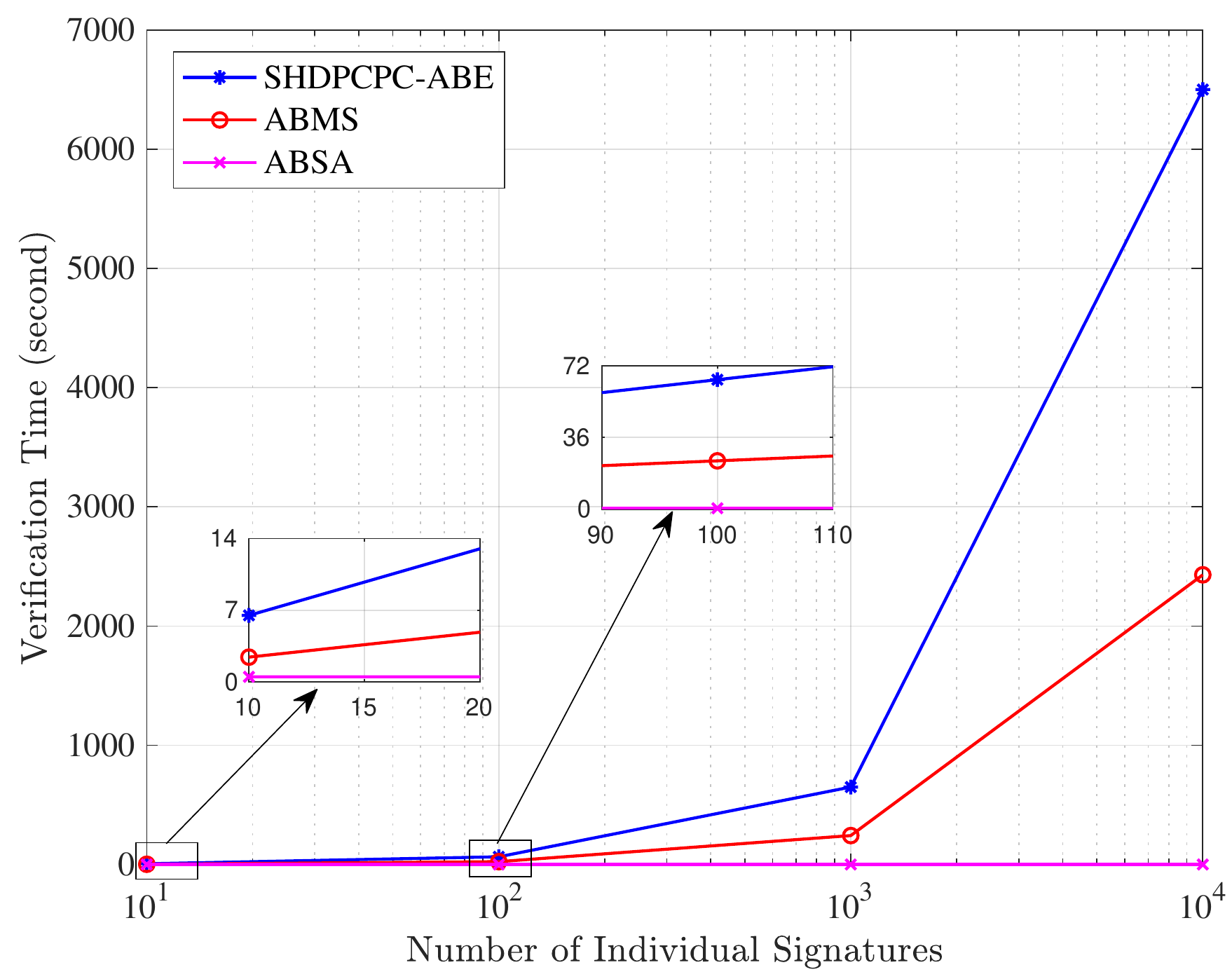}
\caption{Verification time of ABSA, ABMS \cite{guo2020icbc} and SHDPCPC-ABE \cite{li2021ehrchain} vs. the number of individual signatures.}
\label{fig:verification-time}
\end{figure}

In the last, we measure the running time of the ABSA verification phase and compare it with the SHDPCPC-ABE~\cite{xu2019healthchain} and the ABMS~\cite{guo2020icbc} schemes that do not have the aggregation feature in the design. The ring-based AC policy in Fortifiedchain~\cite{egala2021fortified} is not comparable since it does not have cryptographic primitives.  As shown in Fig. \ref{fig:verification-time}, the verification time of the SHDPCPC-ABE and ABMS schemes will increase linearly with increasing the number of individual signatures. Accordingly, the ABSA scheme's verification time will keep constant at 508 ms because multiple individual signatures and verification keys are aggregated to preserve the verification performance and achieve higher efficiency.

\section{Conclusion}
This paper addressed the privacy and security shortcomings in EHR management systems. We proposed a hybrid ({\em on-chain}/{\em off-chain}) architecture of blockchain and edge computing, which incorporates an attribute-based signature aggregation mechanism (ABSA) to authenticate the patients' private attributes and the MA-ABE scheme to secure the EHR access control mechanism. We prototype hybrid architecture using Hyperledger Fabric blockchain, Hyperledger Ursa, and OpenABE cryptographic library. Moreover, We utilize the Paillier homomorphic encryption scheme for EHR ciphertext computation to preserve data owners' privacy. The experiment results show the desirable system performance that meets real-world scenarios' requirements (e.g., response time, network latency) while safeguarding EHR and is robust against unauthorized retrievals.

In the future, we plan to enhance our signature aggregation scheme in the presence of multiple dynamic attributes to provide more flexible access control properties. Also, the hierarchical architecture of a novel blockchain structure to increase network scalability could be another promising research field.

\ifCLASSOPTIONcompsoc
  \section*{Acknowledgments}
\else
  \section*{Acknowledgment}
  This paper is the extended and revised version of \cite{guo2019access} presented at the 2nd IEEE International Conference on Blockchain (Blockchain 2019) and \cite{guo2020icbc} presented at the 2020 IEEE International Conference on Blockchain and Cryptocurrency (ICBC 2020).
  

This work was partially supported by the Fundamental Research Funds for the Central Universities under the Grant G2021KY05101, 2021-2024. This research is supported in part by the Guangdong Basic and Applied Basic Research Foundation under the Grant No. 2021A1515110286, 2021-2024, the Natural Science Foundation of Shaanxi Provincial Department of Education under the Grant No. 2022JQ-639, and a Federal Highway Administration grant: ``Artificial Intelligence Enhanced Integrated Transportation Management System", 2020-2023.
  
\fi

\ifCLASSOPTIONcaptionsoff
  \newpage
\fi

\bibliographystyle{IEEEtran}
\bibliography{sig.bib}


\section*{Biographies}
\vskip -2\baselineskip plus -1fil 

\begin{IEEEbiography}
[{\includegraphics[width=1.0in,height=1.25in,clip]{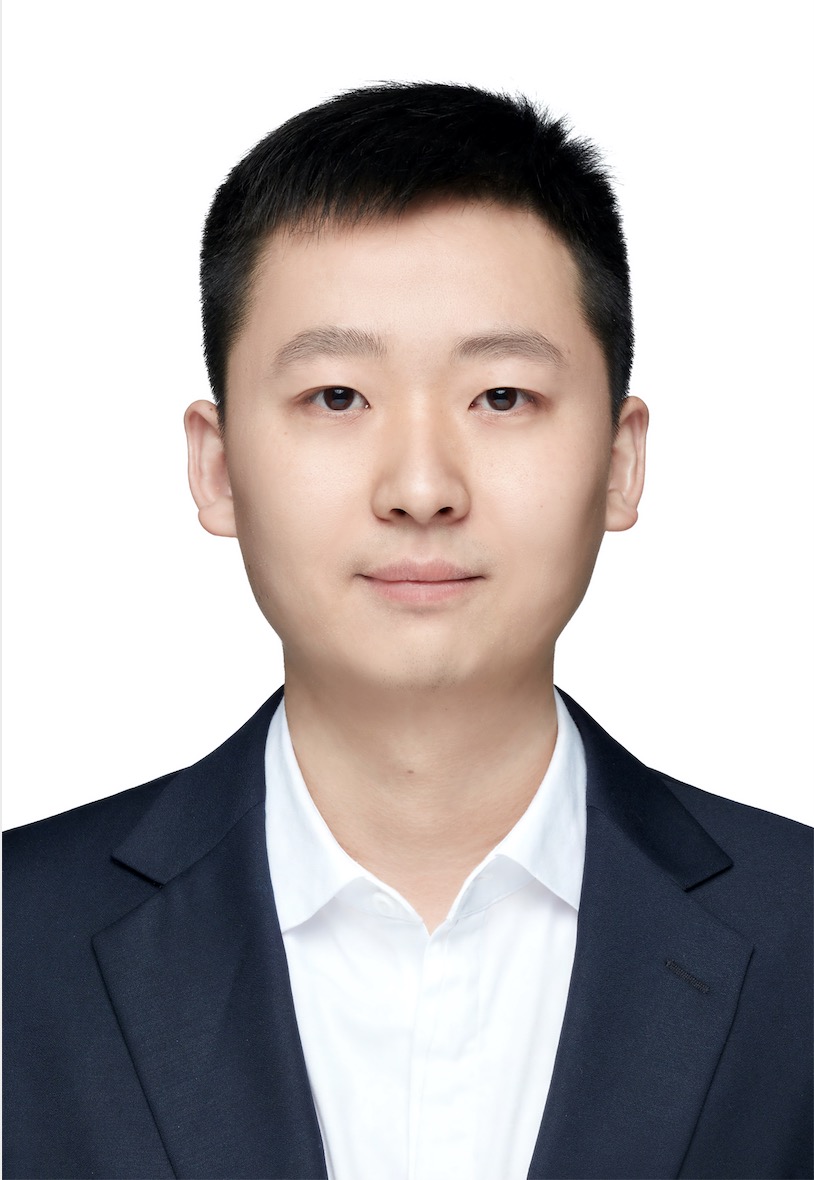}}]{Hao Guo} received the B.S. and M.S. degrees from the Northwest University, Xi'an, China in 2012, and the Illinois Institute of Technology, Chicago, United States in 2014, and his Ph.D. degree from the University of Delaware, Newark, United States in 2020, all in computer science.
He is currently an Assistant Professor
with the School of Software at
the Northwestern Polytechnical University.
His research interests include blockchain and distributed ledger technology, data privacy and security, cybersecurity, cryptography technology, and Internet of Things (IoT). He is a member of both ACM and IEEE.
\end{IEEEbiography}
\vskip -6pt plus -1fil
\begin{IEEEbiography}
[{\includegraphics[width=1in,height=1.25in,clip]{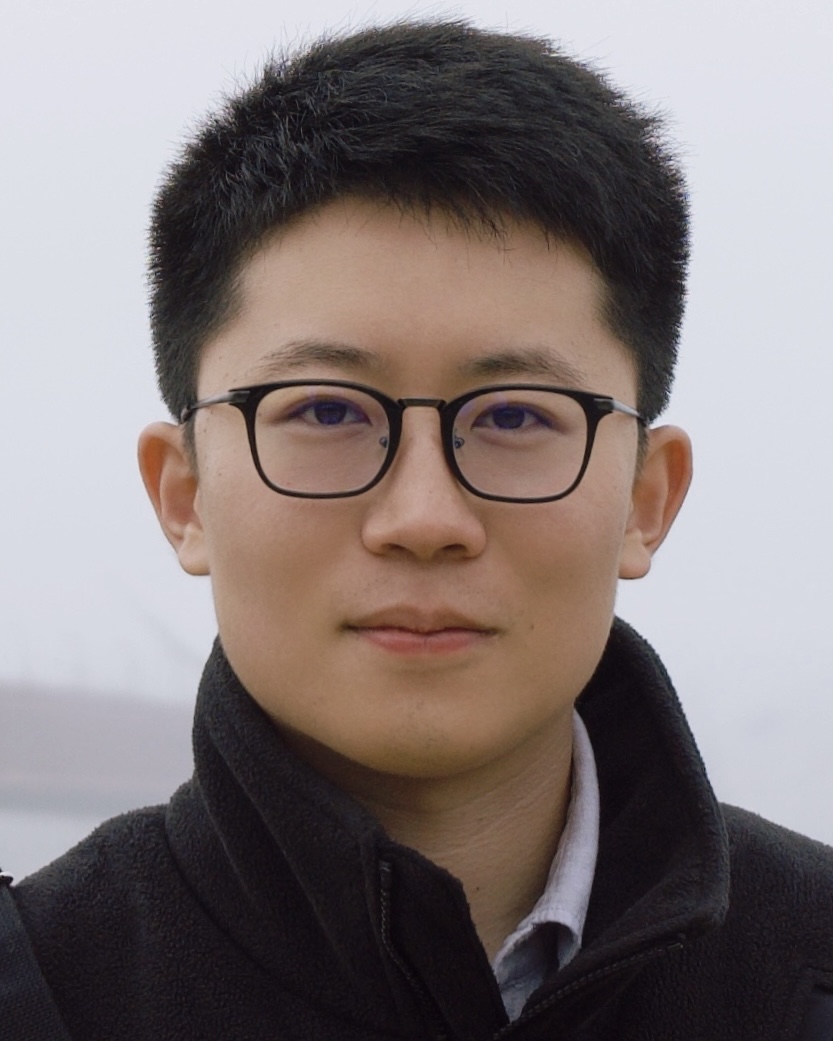}}]{Wanxin Li} (Member, IEEE) received the B.S. degree from the Chongqing University in 2015, and the M.S. and Ph.D. degrees from the University of Delaware in 2017 and 2022, respectively. He is a Lecturer with the Department of Communications and Networking, Xi'an Jiaotong-Liverpool University. His research interests are the privacy and scalability in blockchain, and blockchain-based architecture designs such as connected and autonomous vehicular networks, electronic health records and federated learning. 
\end{IEEEbiography}
\vskip -6pt plus -1fil
\begin{IEEEbiography}
[{\includegraphics[width=1in,height=1.25in,clip]{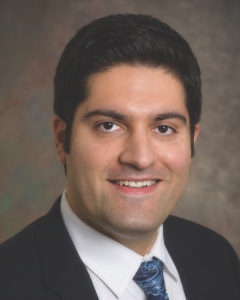}}]{Mark Nejad} is an Assistant Professor at the University of Delaware. His research interests include network optimization, distributed systems, blockchain, game theory, and automated vehicles. He has published more than forty peer-reviewed papers and received several publication awards including the 2016 best doctoral dissertation award of the Institute of Industrial and Systems Engineers (IISE) and the 2019 CAVS best paper award from the IEEE VTS. His research is funded by the National Science Foundation and the Department of Transportation. He is a member of the IEEE and INFORMS.
\end{IEEEbiography}
\vskip -6pt plus -1fil
\begin{IEEEbiography}[{\includegraphics[width=1in,height=1.125in,clip]{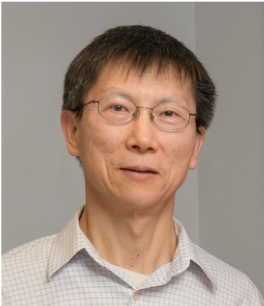}}]{Chien-Chung Shen}
received his B.S. and M.S. degrees from
National Chiao Tung University, Taiwan, and his Ph.D. degree
from UCLA, all in computer science.  He was a research
scientist at Bellcore Applied Research working on control
and management of broadband networks.  He is now a Professor
in the Department of Computer and Information Sciences of
the University of Delaware. His research interests include
blockchain, Wi-Fi, SDN
and NFV, ad hoc and sensor networks, dynamic spectrum management,
cybersecurity, distributed computing, and
simulation.  He is a recipient of NSF CAREER Award and a member of both ACM and IEEE.
\end{IEEEbiography}





\end{document}